\documentclass[%
reprint,
superscriptaddress,
longbibliography,
amsmath,amssymb,
prb,
]{revtex4-2}

\usepackage[version=4]{mhchem}
\setcitestyle{numbers,square}
\usepackage{subfigure}
\usepackage{booktabs}
\usepackage{tabularx}
\usepackage{physics}
\usepackage{graphicx}
\usepackage{dcolumn}
\usepackage{bm}
\usepackage{xcolor}
\usepackage{hyperref}
\usepackage{mathtools}
\usepackage[normalem]{ulem}

\hypersetup{
    colorlinks,%
    citecolor=blue,%
    linkcolor=blue,%
    urlcolor=blue
}

\newcommand{\orcid}[1]{\href{https://orcid.org/#1}{\includegraphics[width=8pt]{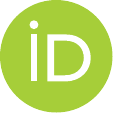}}}

\begin{document}

\title{Tunable Electronic and Transport Properties of Biphenylene via Fluorination and Disorder}
\author{Lucas Soares Sousa \orcid{0000-0002-2685-6531}}
\affiliation{Instituto de F\'isica, Universidade Federal de Uberl\^andia, 38400-902 Uberl\^andia, MG, Brazil}
\author{Felipe Crasto de Lima \orcid{0000-0002-2937-2620}}
\affiliation{Ilum School of Science, Brazilian Center for Research in Energy and Materials (CNPEM), 13083-970 Campinas, S\~ao Paulo, Brazil}
\author{Roberto Hiroki Miwa \orcid{0000-0002-1237-1525}}
\affiliation{Instituto de F\'isica, Universidade Federal de Uberl\^andia, 38400-902 Uberl\^andia, MG, Brazil}
 
\begin{abstract}
Biphenylene (BPN) network is a newly synthesized 2D carbon allotrope hosting anisotropic Dirac electronic states. 
Here, we investigate how fluorination and correlated chemical disorder modify the electronic structure and charge transport of fluorinated biphenylene (F/BPN) using density functional theory, Wannier-based tight-biding Hamiltonian, and quantum transport simulations. 
We show that fluorination reshapes the transport response of BPN, producing concentration-dependent anisotropic conduction regimes. For pristine and ordered fluorinated systems, we identified the emergence of negative differential resistance (NDR) and a bias-induced inversion of the preferred transport direction, from armchair to zigzag and vice versa. In contrast, disorder suppresses the NDR, driving the system toward an approximately Ohmic transport regime. At high fluorine coverage, we further observed a nonmonotonic dependence of the armchair current on adatom concentration, which we attribute to the formation of correlated quasi-linear fluor conformation that promote armchair-oriented C-$\pi$ transport channels while simultaneously suppressing transport along the zigzag direction. Our results demonstrate that correlated fluorination can be used as an active mechanism to engineer electronic transport.

\end{abstract} 

\maketitle

\section{Introduction}

Two-dimensional (2D) carbon allotropes continue to stimulate  research efforts due to their structural diversity and the resulting range of physical and chemical properties \cite{paupitz2026concise, ahmad2024carbon}.
Beyond graphene, several nonbenzenoid 2D carbon networks have been proposed, for instance, T-graphene \cite{liu2012structural}, graphynes and graphdiynes \cite{li2023graphynes, liu2024graphyne}, many of which present electronic behavior distinct from the honeycomb lattice, such as unconventional topology \cite{koizumi2024topological, son2022magnetic, PCCPcrasto2019}.
Among these materials, the biphenylene network (BPN) has recently emerged as a novel planar carbon allotrope composed of fused four-, six-, and eight-membered rings.
The first experimental synthesis of BPN was achieved by on-surface polymerization and subsequent thermal annealing of hexafluoro-dibromo-terphenyl precursors on Au(111), yielding extended atomically thin sheets with high structural order \cite{SCIENCEfan2021}.
The successful realization of BPN opened a path to explore the electronic, mechanical, thermal, and chemical properties of this long-predicted carbon network.

\begin{figure*}
    \centering
    \includegraphics[width=2\columnwidth]{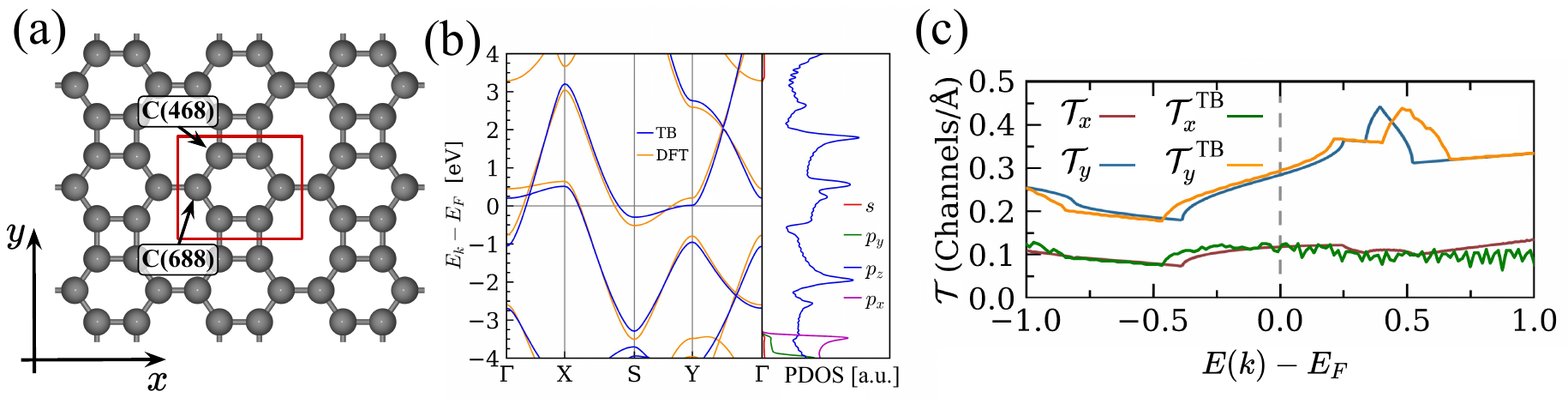}
    \caption{Structural model (a), electronic band structure and the projected density of states (b), and the transmission coefficient (c) of the pristine biphenylene. In (b), the electronic band structure calculated via first-principles density functional theory (DFT) (yellow lines) is superimposed with the optimized Wannier-based tight-binding (TB) model (blue lines), and in (c) the transmission $\mathcal{T}_x$ and $\mathcal{T}_y$ calculated using first-principles DFT are superimposed with those obtained using the TB model, $\mathcal{T}_x^\text{TB}$ and $\mathcal{T}_y^\text{TB}$.}
    \label{fig:bpn}
\end{figure*}

\begin{figure}[h!]
    \centering
    \includegraphics[width=0.5\linewidth]{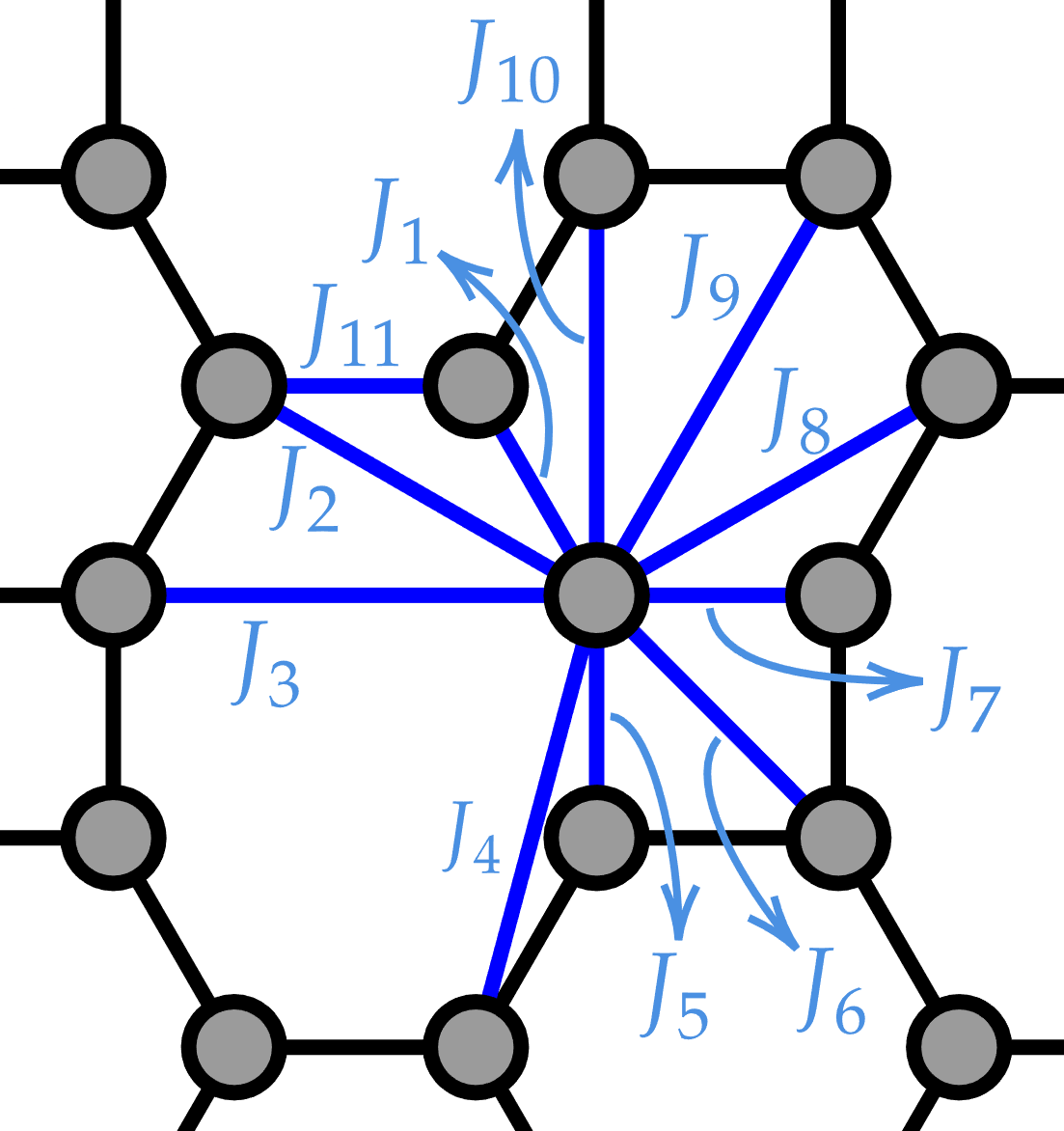}
    \caption{Interaction pairs labels in the disorder model.}
    \label{fig:j_parameters}
\end{figure}

First-principles studies have since revealed that pristine BPN exhibits a rich electronic structure, including strongly anisotropic bands, high electron thermal conductivity~\cite{AEMtong2022}, and type-II Dirac cones protected by the $D_{2h}$ symmetry of the lattice~\cite{PRBliu2021}.
Additional theoretical work has evaluated its mechanical stability~\cite{SRluo2021}, potential for electrocatalysis~\cite{JPCLliu2021}, interfacial properties with metals~\cite{ACSANMlebre2025}, and prospects as an anode material for next-generation batteries~\cite{PCCPhan2022}.
Transport simulations have also indicated a strong directional dependence of the conduction channels~\cite{PCCPkuritza2024}, further highlighting BPN as a promising platform for anisotropic nanoelectronics.

Chemical functionalization offers a powerful strategy to engineer the electronic structure of 2D materials.
Fluorination in particular has proven effective in modulating band gaps, localization effects, and transport properties in graphene-derived systems~\cite{ASfeng2016}.
Recent theoretical work has examined fluorinated variants of BPN, predicting that fluorine adsorption can break key lattice symmetries, modify Dirac's fermion features, alter charge distribution, and lead to tunable electronic phases~\cite{NLmo2024}.
However, a systematic understanding of the role of edges, finite-size effects, disorder induced by partial functionalization, or the emergence of transport phenomena remains incomplete.
Because the biphenylene lattice is intrinsically anisotropic, chemical functionalization and correlated disorder may couple strongly to transport directionality, offering a route to tailor transport anisotropy.
Beyond conventional band engineering, controlled disorder \cite{NRCsimonov2020} has recently emerged as a route to induce unconventional transport regimes in low-dimensional materials \cite{PRLdassarma2011, NATCOMMcho2018, COMMPHYSneverov2022, AMTdasneves2025}.
Contrary to uncorrelated Anderson-like disorder, which leads to localization \cite{PRanderson1958}, in correlated disordered systems, impurity interactions may reorganize electronic pathways, generate anisotropic localization effects \cite{PRLizrailev1999, PRLbodyfelt2014, PEtolozaSandoval2025}.

In this work, we investigate how fluorination and correlated disorder reshape the electronic and transport properties of biphenylene networks by combining density functional theory (DFT), Wannier-based tight-binding Hamiltonians, and quantum transport simulations. We showed that fluorination not only modifies the Dirac electronic structure of biphenylene but also gives rise to tunable anisotropic transport regimes that depend sensitively on the concentration and spatial distribution of fluorine adatoms.
Our results demonstrate that correlated chemical disorder can be used as an active mechanism to engineer emergent transport phenomena and directional current control in BPN-based nanoelectronic systems. 

\section{Computational Details}

\subsection{Ab initio calculations}

Our density functional theory calculations were performed from first principles using the VASP (Vienna Ab initio Simulation Package) \cite{PRBkresse1996}.
The exchange-correlation interaction is treated within the Generalized Gradient Approximation (GGA) using the Perdew-Burke-Ernzerhof (PBE) exchange-correlation\cite{PRBperdew1996} and van der Waals (vdW) corrections are incorporated via the DFT-D3 method \cite{JCPgrimme2010}.
The ionic cores were described using the Projector Augmented Wave (PAW) method \cite{PRBblochl1994, PRBkresse1999}, while the electronic wave functions were expanded in a plane-wave basis set with a kinetic energy cutoff of $550$\,eV, ensuring convergence of the total energy.
For Brillouin zone integration, k-point meshes of $6 \times 6 \times 1$ were used for structural relaxation and a dense $\Gamma$-centered k-point mesh of $24\times24\times1$ for electronic structure calculations.
Structural optimization continued until the energy difference dropped below $10^{-6}$\,eV and the forces on each atom were less than $10^{-3}$\,eV/{\AA}.

\subsection{Effective tight-binding model}

Although DFT provides a realistic description of the electronic structure, its computational cost scales approximately as $O(N^3)$ with the number of atoms, making simulations of disordered ribbons containing thousands of atoms computationally impractical.
To overcome this limitation, we constructed an effective tight-binding Hamiltonian based on Maximally Localized Wannier Functions (MLWFs) using the Wannier90 package \cite{JPCMpizzi2020}, with the Hamiltonian
\begin{equation}
       H = \sum_{i} \epsilon_i c_i^\dagger c_i + \sum_{\langle i,j \rangle} t_{ij} (c_i^\dagger c_j + \text{h.c.}),
\end{equation}
where $\epsilon_i$ represents the on-site energy and $t_{ij}$ denotes the hopping integrals. Figure~\ref{fig:bpn}(a) presents the structural model of pristine BPN. The orbital-projected density of states (PDOS), shown in Fig.~\ref{fig:bpn}(b), reveals that the electronic states near the Fermi level are dominated by carbon $p_z$ orbitals.
This observation justifies the construction of a Wannier-based minimal tight-binding model containing a single effective orbital per site.
Although the resulting MLWFs are not strictly identical to atomic $p_z$ orbitals, they retain the same dominant orbital character and accurately reproduce the low-energy band structure within the relevant energy window. For the fluorinated systems, the dominance of $p_z$ orbitals near the Fermi level causes $sp^3$ hybridization of the C and F atoms be far from the Fermi level, so that in our TB model the fluorinated C atoms as like the F atoms was modeled as a vacancy.

\subsection{Transport calculations}

To accurately characterize the charge transport dynamics across diverse length scales, two formally equivalent computational frameworks are utilized: (i) first-principles calculations using the non-equilibrium Green's function (NEGF)~\cite{keldysh2024diagram, haug2008quantum} formalism implemented with TranSIESTA package \cite{PRBbrandbyge2002}, and (ii) to simulate very large systems we perform complementary tight-binding simulations~\footnote{Our TB approach does not solve the Keldysh nor the Poisson equations self-consistently, consequently, neglecting non-equilibrium charge redistribution. In order to get some insight about this configuration, we employ a linear potential drop in the scattering region matching the bias values at the electrodes} evaluating the scattering matrix directly using the Kwant~\cite{groth2014kwant} package.
Both methodologies operate within the coherent steady-state transport regime described by the Landauer-Büttiker formalism, where the linear-response conductance $G$ at zero temperature is governed by the energy-dependent transmission probability $\mathcal{T}(E)$, such that $G = (2e^2/h) \mathcal{T}(E_F)$, where $E_F$ is the Fermi energy \cite{datta1997electronic}.
The device is modeled as a central scattering region connected to two semi-infinite electrodes (Left and Right leads). The macroscopic steady-state current [$I(V)$] is calculated using the Landauer–Büttiker formula
\begin{equation}
    I(V) = \frac{2e}{h}\int_{-\infty}^{+\infty}\dd{E}\mathcal{T}(E,V)\bqty{f(E-\mu_L) - f(E-\mu_R)},\\
\end{equation}
where $f(E-\mu_{L/R}) = \bqty{e^{\pqty{E-\mu_{L/R}}/k_B T} + 1}^{-1}$, and $\mu_{L/R}$ are the chemical potentials of the left/right leads for a given voltage $V$ \cite{datta1997electronic}.

\subsection{Disordered distribution}

To model the stochastic nature of fluorine adsorption, we employ a correlated disorder model. The distribution of fluorine adatoms is not random but governed by thermodynamic interactions between dopants.
We use a Metropolis-Hastings Monte Carlo algorithm to generate the F/BPN configurations with the configuration energy
 \begin{equation}
     E_{\text{config}} = -\sum_{\langle i,j \rangle}J_{ij}\sigma_{ij} + \sum_{i}E_i^B, \label{B1}
 \end{equation}
where the first sum runs for all pairs $\langle i,j\rangle$ within the pair distance threshold $3.5$\,{\AA}.
The variable $\sigma_{ij}$ equals $1$ if both adjacent lattice sites are occupied with fluor, and $0$ otherwise. $E^B_i$ is the fluor binding energy in the $i$-th carbon atom.
$J_{ij}$ is the inter-fluor interaction energy.
Using DFT calculations, we can compute the interaction energy
 \begin{equation}
     J_{ij} = (E^{ij} + E^0) - (E_i + E_j)
 \end{equation}
where $E^{ij}$ is the energy of the system with the two impurities, $E^0$ is the energy of the pristine system, $E_i$ and $E_j$ are the energy of the system in a single impurity $i$ and $j$, respectively.
The pairs and values of $J_{ij}$ are shown in Fig.~\ref{fig:j_parameters} and Table~\ref{tab:j_parameters} respectively.

The disorder is applied only in the scattering region, while the semi-infinite electrodes are maintained with a perfectly ordered, periodic fluorine concentration.
We have used $2\times 10^5$ equilibrium steps for the final configuration, simulated with a temperature scale $k_BT =25$\,meV, and the transport analysis was done by averaging $64$ independent final disorder distributions.

\begin{table}[h!]
\caption{\label{tab:placeholder} \label{tab:j_parameters} Calculated F adsorption model pair site distance (in \AA), inter-fluor interaction energy $J_{ij}$ (in eV), and binding energy $E_i^B$ (in eV) for each labels described in the Fig.~\ref{fig:j_parameters}.}
\begin{ruledtabular}
    \begin{tabular}{ccr|cc}
        Coupling & Distance [\AA]& $J_{ij}$ & Site & $E_i^B$ \\
        \hline
$J_{1}$  & 1.41  & -0.175 & C(468) & 3.20 \\
$J_{2}$  & 2.53  & 0.197  & C(688) & 2.21 \\
$J_{3}$  & 3.06  & -0.080 & & \\
$J_{4}$  & 3.45  & 0.220  & & \\
$J_{5}$  & 1.46  & 0.278  & & \\
$J_{6}$  & 2.06  & 0.986  & & \\
$J_{7}$  & 1.45  & -0.522 & & \\
$J_{8}$  & 2.54  & 0.100  & & \\
$J_{9}$  & 2.72  & -0.251 & & \\
$J_{10}$ & 2.30  & 0.861  & & \\
$J_{11}$ & 1.45  & 0.289  & & \\
    \end{tabular}
\end{ruledtabular}
\end{table}

\begin{figure}
    \centering
    \includegraphics[width=\linewidth]{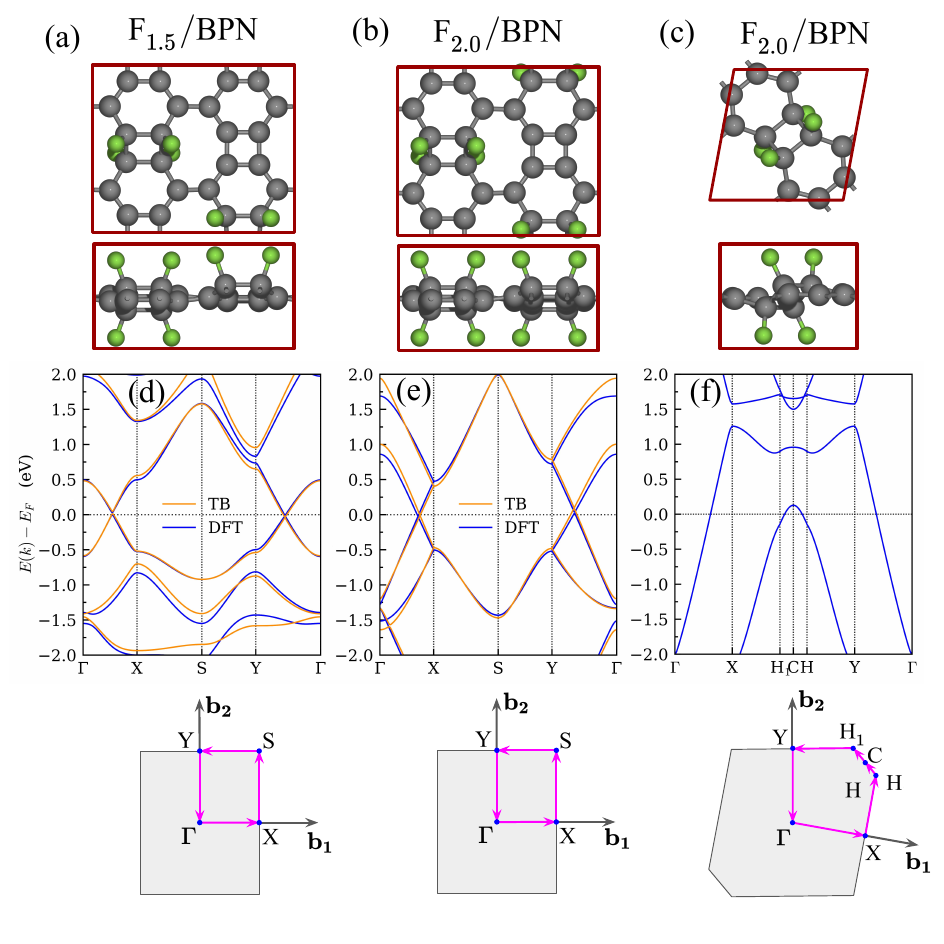}
    \caption{Structural models of F$_{1.5}$/BPN (a), F$_{2.0}$/BPN (b), the primitive unit cell of F$_{2.0}$/BPN (c), and the respective electronic band structures and Brillouin zones (d)-(f) indicating the high symmetry points.}
    \label{fig:bands}
\end{figure}

\section{Results and Discussions}

\begin{figure}
    \centering
    \includegraphics[width=\linewidth]{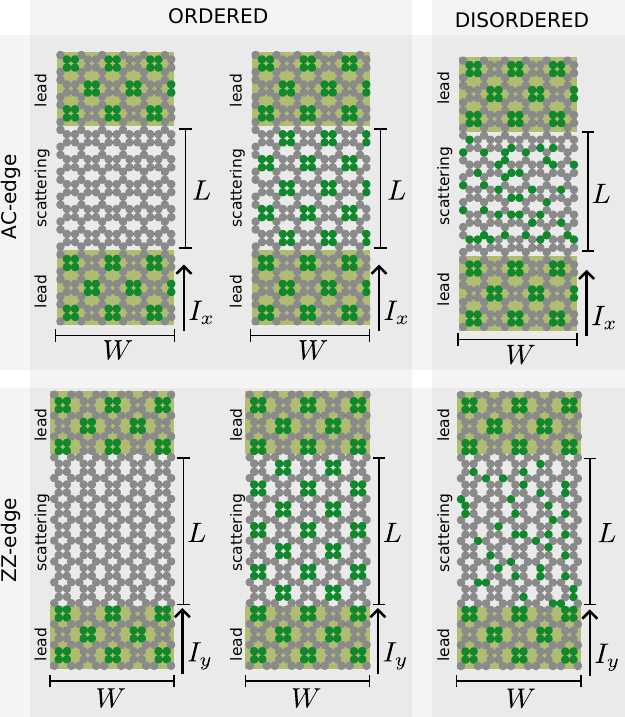}
    \caption{Schematic setup of two-terminal devices,  pristine BPN, ordered, and disordered F/BPN scatterings connected to F$_n$/BPN leads ($n$=1.5 and 2.0), with scattering lengths of $L=4.5$\,nm (AC-edge) and $3.8$\,nm (ZZ-edge) for the pristine and ordered systems, and  $L=9.0$\,nm (AC-edge) and $7.5$\,nm (ZZ-edge) for disordered F/BPN. The TB calculations were performed using nanoribbons with finite widths $W$,  while periodic boundary conditions (perpendicularly to the electronic transport direction) were used in the DFT calculations.  \label{fig:edges_definition}}
\end{figure}

\subsection{Pristine and Fluorinated Biphenylene}

Biphenylene (BPN) is a two-dimensional carbon allotrope that crystallizes in a planar $D_{2h}$ structure.
This lattice is constructed from the periodic fusion of 4-, 6-, and 8-membered rings, resulting in a rectangular Bravais lattice with lattice constants $a (\hat x)$ and $b (\hat y)$ [Fig.~\ref{fig:bpn}(a)] of $4.51$ and $3.76$\,{\AA}, respectively, in agreement with the values reported experimentally, $4.52$ and $3.76$\,{\AA} ~\cite{SCIENCEfan2021}. BPN exhibits two distinct sublattices defined by their local ring environments: four C$(468)$ sites, located at the vertices shared by a square, a hexagon, and an octagon, and two C$(688)$ sites, located at the vertices shared by one hexagon and two octagons.
The electronic band structure of pristine BPN [Fig.~\ref{fig:bpn}(b)] exhibits intrinsic metallic character, with the C-$2p_\text{z}$ orbitals of the C$(468)$ sites providing the dominant contribution to the formation of the metallic bands \cite{PCCPkuritza2024}, which are characterized by the emergence of type-II Dirac cones \cite{PRBliu2021}.
Their stability is robustly protected by crystalline mirror symmetry and $C_2$ rotational symmetry.
It is noteworthy that further theoretical analysis indicates that these symmetries enforce the existence of symmetry-related ``partner'' Dirac points; for each Dirac point at momentum $k$, there exists a corresponding point at $-k$ or at a symmetry-equivalent position. Consequently, the number of Dirac points in the full Brillouin zone is twice that typically reported.

Chemical functionalization of BPN via adsorption of foreign species, particularly hydrogen and halogen atoms~\cite{xie2022effective}, is an effective strategy for engineering its electronic structure.
As shown by Mo et al.~\cite{NLmo2024}, controlled fluorination of biphenylene (F/BPN) induces both type-I and type-II Dirac cones.  Figure~\ref{fig:bands} presents F/BPN, and the corresponding electronic band dispersions; here, the concentration of F adatoms is defined per formula unit of the C$_6$ BPN. Fluorine adsorption acts as a symmetry-breaking perturbation that may preserve or break the $M_y$ mirror symmetry protecting the Dirac crossings.
In agreement with Ref.~\cite{NLmo2024}, we obtain type-II Dirac cones for F$_{0.5}$/ and F$_{1.0}$/BPN (not shown), and type-I (massive and massless) Dirac points in F$_{1.5}$/BPN, Fig.~\ref{fig:bands}(d).
However, the previously reported Dirac nodal line in F$_{2.0}$/BPN [Fig.~\ref{fig:bands}(e)] is a spurious artifact arising from band folding, since the system actually possesses a smaller unit cell, Fig~\ref{fig:bands}(c).
As a result, the electronic structure is characterized by a metallic band and a hole pocket, as shown in the Fig.~\ref{fig:bands}(f).

\subsection{Electronic transport}

The emergence of isolated type-I Dirac points in F$_{1.5}$/BPN and the contrasting metallic dispersion of F$_{2.0}$/BPN near the Fermi level make these systems promising candidates for two-dimensional electronic devices with highly  tunable electronic transport properties. Motivated by these features, we investigate electronic transport in {\it (1)} pristine BPN, and fluorinated BPN forming {\it (2)} ordered (F$^\text{o}$/BPN) and {\it (3)} disordered (F$^\text{d}$/BPN) structures of F adatoms on BPN  connected to F$_{1.5}$/BPN and F$_{2.0}$/BPN leads. These systems are  schematically illustrated in Fig.~\ref{fig:edges_definition}, and (here) labeled as F$_n$/BPN/F$_n$, F$_n$/F$^\text{o}_n$/F$_n$, and F$_n$/F$^\text{d}_{x_\text{ad}}$/F$_n$, respectively, with $n$=1.5 and 2.0. 

Electronic transport calculations were performed using two approaches: (i) first-principles density functional theory (DFT) calculations and (ii) an effective single-orbital tight-binding (TB) Hamiltonian, both described in Sec.~II. 
This approximation is computationally efficient and has been validated against full DFT calculations. In Figs.~\ref{fig:bpn}(b) and (c), we compare the DFT and TB results for the electronic band structure and the electronic transmission of pristine BPN, and in Figs.~\ref{fig:bands}(d) and (e) we compare the DFT and TB electronic band structures of the fluorinated systems, F$_{1.5}$/ and F$_{2.0}$/BPN.

\subsubsection{Pristine scatterer: {\rm F$_n$/BPN/F$_n$}}

The metallic character of pristine BPN is primarily governed by the carbon $2p_z$ orbitals associated with the square lattice [C(468) in Fig.~\ref{fig:bpn}(a)]. Combined with the anisotropic arrangement of these square lattices, this electronic structure gives rise to a pronounced directional dependence of the transport properties in BPN. Indeed, for a pristine BPN scattering region connected to pristine BPN leads (BPN/BPN/BPN), we find a higher electronic transmittance along the zigzag (ZZ) direction than along the armchair (AC) direction, i.e., $\mathcal{T}_y > \mathcal{T}_x$ [Fig.~\ref{fig:bpn}(c)]. In addition, the current along the AC direction ($I_x$) exhibits negative differential resistance (NDR) (not shown), consistent with the results reported in Ref.~\cite{PCCPkuritza2024}.

We next examine the role of fluorinated leads in the electronic transport through pristine BPN. Our DFT results for the current in F$_n$/BPN/F$_n$, shown as solid lines in Figs.~\ref{fig:current_landauer}(a) and (b), reveal that the F$_{1.5}$/BPN/F$_{1.5}$ junction exhibits lower current values than the corresponding F$_{2.0}$/BPN/F$_{2.0}$ system. Nevertheless, both systems display negative differential resistance along the armchair direction ($I_x$), albeit at different threshold voltages, namely $\Delta V \gtrsim 0.5$ V for F$_{1.5}$/BPN/F$_{1.5}$ and $\Delta V \gtrsim 0.7$ V for F$_{2.0}$/BPN/F$_{2.0}$.

In contrast, the current along the ZZ direction ($I_y$) exhibits a nearly Ohmic behavior for both types of fluorinated leads. However, unlike the pristine BPN/BPN/BPN system, the AC current dominates at low bias voltages. For instance, in F$_{1.5}$/BPN/F$_{1.5}$ [ solid lines in Fig.~\ref{fig:current_landauer}(a)], we find $I_x>I_y$ for $\Delta V \lesssim 0.7$ V, whereas $I_y>I_x$ for $\Delta V \gtrsim 0.7$ V.

Defining the current anisotropy factor as
\begin{equation}
\eta=\frac{I_y}{I_x},
\end{equation}
we obtain $\eta<1$ for $\Delta V \lesssim 0.7$ V and $\eta>1$ for $\Delta V \gtrsim 0.7$ V.
Notably, a similar voltage-induced inversion of the transport anisotropy is also observed in F$_{2.0}$/BPN/F$_{2.0}$, although with distinct characteristics, as shown in Fig.~\ref{fig:current_landauer}(c).

The electronic transport calculations based on the effective tight-binding Hamiltonian~\cite{groth2014kwant} were performed using ribbon structures with widths of $W=38$ and $45$\,nm and the same scattering lengths, $L$, adopted in the DFT calculations.
Notably, the tight-binding results, shown as dashed lines in Figs.~\ref{fig:current_landauer}(a) and (b), successfully reproduce the main transport features predicted by DFT, including the lower current values in F$_{1.5}$/BPN/F$_{1.5}$, the emergence of NDR in $I_x$, and the nearly Ohmic behavior of $I_y$.

\subsubsection{Ordered scatterer: {\rm \rm F$_n$/F$^\text{o}_n$/F$_n$}}

Here, we examine the electronic transport in ordered systems, F$_n$/F$^\text{o}_n$/F$_n$, in which the geometry of the leads and the scattering regions is the same, Fig.~\ref{fig:edges_definition}.

\begin{figure}
    \centering
    \includegraphics[width=\columnwidth]{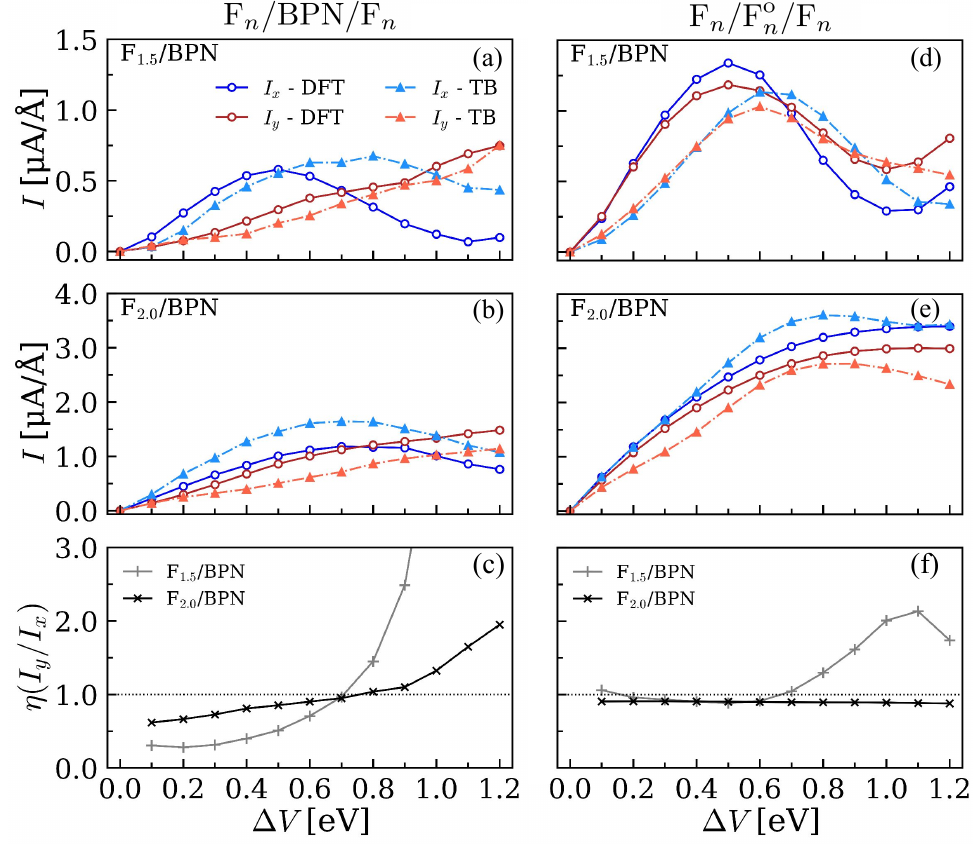}
    \caption{Current as a function of bias voltage for pristine BPN connected to F$_{1.5}$/BPN (a) and F$_{2.0}$/BPN (b) leads, and the corresponding current anisotropy factor $\eta$ (c). Current through ordered fluorinated BPN, F$_{1.5}$/F$^\text{o}_{1.5}$/F$_{1.5}$ (d) and F$_{2.0}$/F$^\text{o}_{2.0}$/F$_{2.0}$ (e), together with the corresponding current anisotropy factor (f).}
    \label{fig:current_landauer}
\end{figure}

Our results for the current [$I(V)$] in F$_{1.5}$/F$^\text{o}_{1.5}$/F$_{1.5}$ and F$_{2.0}$/F$^\text{o}_{2.0}$/F$_{2.0}$, based on first-principles DFT calculations [solid lines in Figs.~\ref{fig:current_landauer}(d) and (e), respectively], reveal a strong dependence of the $I(V)$ characteristics on the concentration of fluorine adatoms on BPN, and a reduction in the anisotropic features of the current when compared with that of pristine BPN systems, namely BPN/BPN/BPN \cite{PCCPkuritza2024} and F$_n$/BPN/F$_n$. 

The current in F$_{1.5}$/F$^\text{o}_{1.5}$/F$_{1.5}$ is characterized by the emergence of negative differential resistance (NDR) along both transport directions, namely armchair (AC) and zigzag (ZZ).
Specifically, the current initially increases with bias voltage up to a threshold value of $V_\text{th}\approx0.5$ V.
Beyond this threshold, (i) the current decreases markedly, signaling the onset of NDR, and (ii) the transport anisotropy is inverted, such that the current along the ZZ direction becomes larger than that along the AC direction.
That is, we find a transition from $\eta<1$ to $\eta>1$ for $V>V_\text{th}$ in  F$_{1.5}$/F$^\text{o}_{1.5}$/F$_{1.5}$ [Fig.~\ref{fig:current_landauer}(f)]. 

In contrast, (iii) no NDR is observed in F$_{2.0}$/F$^\text{o}_{2.0}$/F$_{2.0}$. Instead, the $I$--$V$ characteristics [Fig.~\ref{fig:current_landauer}(e)] exhibit current saturation for $V\gtrsim1.2$ V, accompanied by an almost bias-independent anisotropy factor of $\eta\approx0.9$ [Fig.~\ref{fig:current_landauer}(c)]; and remarkably, (iv) despite the higher fluorine-adatom concentration, F$_{2.0}$/F$^\text{o}_{2.0}$/F$_{2.0}$  exhibits larger current values compared with its counterpart F$_{1.5}$/F$^\text{o}_{1.5}$/F$_{1.5}$.

A comparison between the two fluorination levels further reveals that the transport anisotropy is significantly enhanced in F$_{1.5}$/F$^\text{o}_{1.5}$/F$_{1.5}$ for bias voltages above $0.8$ V.
This behavior can be associated with the pronounced anisotropy of the electronic dispersion along the $\Gamma-\text{X}$ and $\Gamma-\text{Y}$ directions in F$_{1.5}$/BPN, Fig.~\ref{fig:bands}(d). In contrast, the more symmetric band dispersion in F$_{2.0}$/BPN [Fig.~\ref{fig:bands}(e)] is consistent with its weaker transport anisotropy.

The calculations of the electronic transport properties based on the effective TB Hamiltonian~\cite{groth2014kwant} were performed by using ribbon structures, with ribbon widths, $W$, of $38$ and $45$\,nm, and the same scattering lengths, $L$, as used in the DFT calculations.
Our TB results, shown as dashed lines in Fig.~\ref{fig:current_landauer}, reveal that the features (i)–(iv) discussed above, obtained from first-principles DFT calculations, are well captured by the vacancy approach used to describe the C–F bonds.
In particular, it is quite remarkable that the present TB approach successfully capture both the emergence of the NDR and the bias-induced anisotropy inversion predicted by the first-principles DFT-NEGF calculations in F$_{1.5}$/BPN.
This is an important result, as we will subsequently discuss the effects of disorder on electronic transport in F/BPN using the effective TB approach.

\subsubsection{Disordered scatterer: {\rm \rm F$_n$/F$^\text{d}_{x_\text{ad}}$/F$_n$}}

The introduction of structural disorder or random impurity scattering may drive the system into the regime of Anderson localization.
In this limit, itinerant electrons undergo multiple scattering events with random phase accumulation, leading to destructive interference that exponentially suppresses wavefunction propagation and significantly reduces the localization length \cite{cresti2008charge, mucciolo2009conductance}.

Here, we consider disorder effects within the scattering region, arising from a random distribution of fluorine adatoms, with concentration $x_\text{ad}$, coupled to ordered leads, forming F$_{1.5}$/F$^\text{d}_{x_\text{ad}}$/F$_{1.5}$ and F$_{2.0}$/F$^\text{d}_{x_\text{ad}}$/F$_{2.0}$ junctions, as schematically illustrated in Fig.~\ref{fig:edges_definition} with nanoribbon widths $W$ of $23$\,nm (AC) and $27$\,nm (ZZ).
The spatial distribution of the adatoms is generated through a stochastic process governed by pairwise interactions ($J_{ij}$) and on-site binding energies ($E_i^B$) defined in the Methods section, Eq.~\eqref{B1}.

\begin{figure}
    \centering
    \includegraphics[width=1\columnwidth]{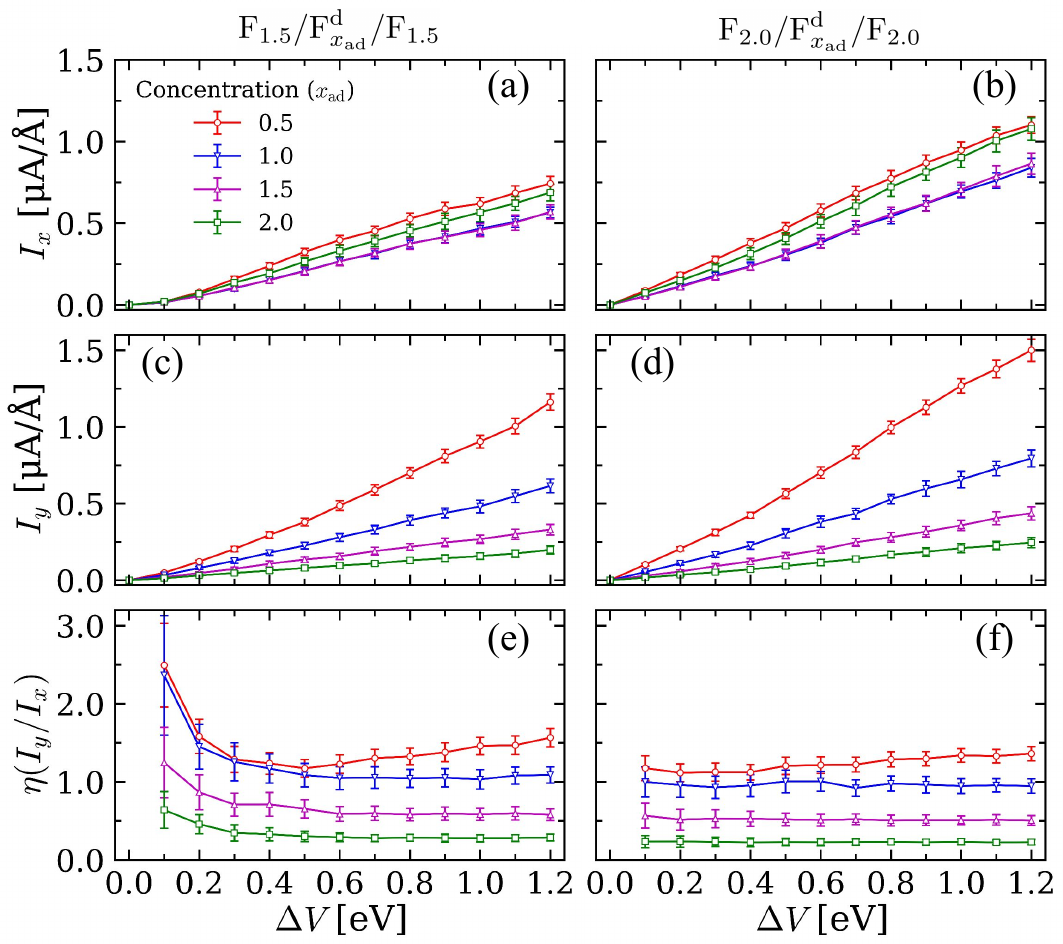}
    \caption{Current as a function of the applied voltage and concentration of fluorine adatoms ($x_\text{ad}$) in disordered fluorinated BPN along the armchair direction ($I_x$) connected to F$_{1.5}$/BPN (a) and F$_{2.0}$/BPN (b) leads.   Current in disordered fluorinated BPN along the zigzag direction ($I_y$) connected to F$_{1.5}$/BPN (c) and F$_{2.0}$/BPN (d) leads; current anisotropy factor in F$_{1.5}$/F$^\text{d}_{x_\text{ad}}$/F$_{1.5}$ (e) and F$_{2.0}$/F$^\text{d}_{x_\text{ad}}$/F$_{2.0}$ (f). }
        \label{fig:js}
\end{figure}

The current as a function of applied bias and fluorine concentration, $I(V,x_\text{ad})$, presented in Fig.~\ref{fig:js}, is obtained by averaging over 64 independent disordered configurations for each value of $x_\text{ad}$.
We find that the inclusion of disorder for $x_\text{ad} \geq 0.5$ leads to a nearly Ohmic transport regime, independent of the transport direction (armchair or zigzag).
Notably, the negative differential resistance (NDR), for $I_x$ and $I_y$, observed in ordered F$_{1.5}$/F$^\text{o}_{1.5}$/F$_{1.5}$ [Fig.~\ref{fig:current_landauer}(d)], is fully suppressed upon the introduction of disorder in F$_{1.5}$/F$^\text{d}_{1.5}$/F$_{1.5}$ [purple solid line in Figs.~\ref{fig:js}(a) and (c)]. Here, the emergence of quasi-Ohmic behavior and the suppression of NDR can be understood in terms of disorder-induced decoherence of charge carriers, which results in a finite lifetime of the electronic states due to the associated energy broadening~\cite{datta1997electronic}.

\begin{figure}
    \centering
    \includegraphics[width=\linewidth]{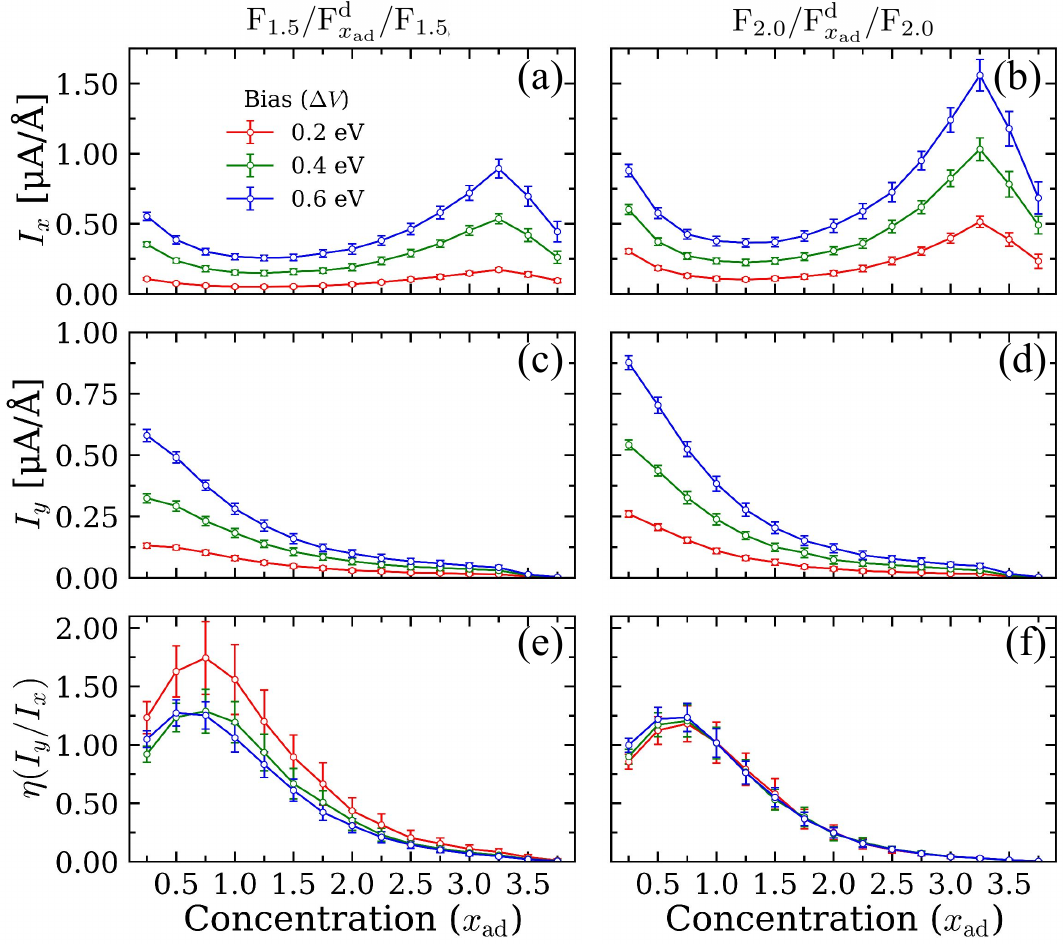}
    \caption{Current as a function of the concentration of fluorine adatoms ($x_\text{ad}$) and bias voltage ($\Delta V$) in disordered fluorinated BPN along the armchair direction ($I_x$) connected to F$_{1.5}$/BPN (a) and F$_{2.0}$/BPN (b) leads. Current in disordered fluorinated BPN along the zigzag direction ($I_y$) connected to F$_{1.5}$/BPN (c) and F$_{2.0}$/BPN (d) leads;   current anisotropy factor in F$_{1.5}$/F$^\text{d}_{x_\text{ad}}$/F$_{1.5}$ (e) and F$_{2.0}$/F$^\text{d}_{x_\text{ad}}$/F$_{2.0}$ (f). }
    \label{fig:rand_currents}
\end{figure}

In this case, for both transport directions, the nearly Ohmic behavior can be expressed as, 
\begin{equation}
V_x = R_x(x_{\rm ad}) I_x,
\qquad
V_y = R_y(x_{\rm ad}) I_y,
\label{eq:resist}
\end{equation}
where the effective resistance $R_i(x_{\rm ad})$ are nearly independent of the applied bias, particularly for  $\Delta V > 0.3$\,V. 
Consequently, the anisotropy factor, $\eta$,
also becomes nearly bias independent. On the other hand, $\eta$ exhibits a noticeable dependence on the fluorine adatom concentration.
As shown in Figs.~\ref{fig:js}(e) and (f), we find that $\eta$ changes from $\eta \geq 1$ to $\eta < 1$ as the fluorine adatom concentration increases from $x_\text{ad}=0.5$ to $2.0$.
In other words, at low fluorine adatom coverage ($x_\text{ad} \leq 1.0$), the current along the zigzag direction is larger than that along the armchair direction, i.e., $I_y>I_x$. In contrast, for $x_\text{ad}>1.0$, we find that $I_y<I_x$ ($\eta<1$). 

In addition to this directional anisotropy, the current along the zigzag direction decreases monotonically with increasing fluorine adatom concentration [Figs.~\ref{fig:js}(c) and (d)].
By contrast, $I_x$ exhibits lower current values at intermediate fluorine concentrations ($x_\text{ad}=1.0$ and $1.5$) and higher values at the extreme concentrations ($x_\text{ad}=0.5$ and $2.0$) [Figs.~\ref{fig:js}(a) and (b)].
To gain further insight into the relationship between the current and the fluorine adatom concentration, $x_\text{ad}$, we analyze the current $I(V,x_{\mathrm{ad}})$ as a function of $x_{\mathrm{ad}}$ within the range $0.25 \leq x_{\mathrm{ad}} \leq 3.75$, for fixed bias voltages of $V = 0.2$, $0.4$, and $0.6$ V. Figures~\ref{fig:rand_currents}(a)-(b) and (c)-(d) show the corresponding results for $I_x(V,x_{\mathrm{ad}})$ and $I_y(V,x_{\mathrm{ad}})$, respectively.
We find that the current along the zigzag direction, $I_y$, decreases asymptotically toward zero with increasing fluorine adatom concentration, independently of the applied bias voltage [Figs.~\ref{fig:rand_currents}(c) and (d)]. In terms of Eq.~\ref{eq:resist}, $R_y(x_\text{ad})$ increases with fluorine adatom concentration, as expected from the increasing number of scattering centers.

In contrast, as shown in Figs.~\ref{fig:rand_currents}(a) and (b), the current along the armchair direction exhibits a nonmonotonic behavior. Independently of the applied bias voltage, $I_x$ initially decreases with increasing fluorine adatom concentration, reaching a minimum around $x_\text{ad}\approx1.25$, and subsequently increases up to a maximum at $x_\text{ad}=3.25$. 
Focusing on the current directional anisotropy, our results of $\eta$, summarized in Figs.~\ref{fig:rand_currents}(e) and (f), not only confirm the predominance of $I_y$ ($I_x$) over $I_x$ ($I_y$) for $x_\text{ad}\leq1$ ($>1$), but also reveal that the behavior of $\eta$ depends on the fluorine adatom concentration in the electrodes, namely, F$_{1.5}$/BPN and F$_{2.0}$/BPN.

\begin{figure}
    \centering
    \includegraphics[width=0.9\columnwidth]{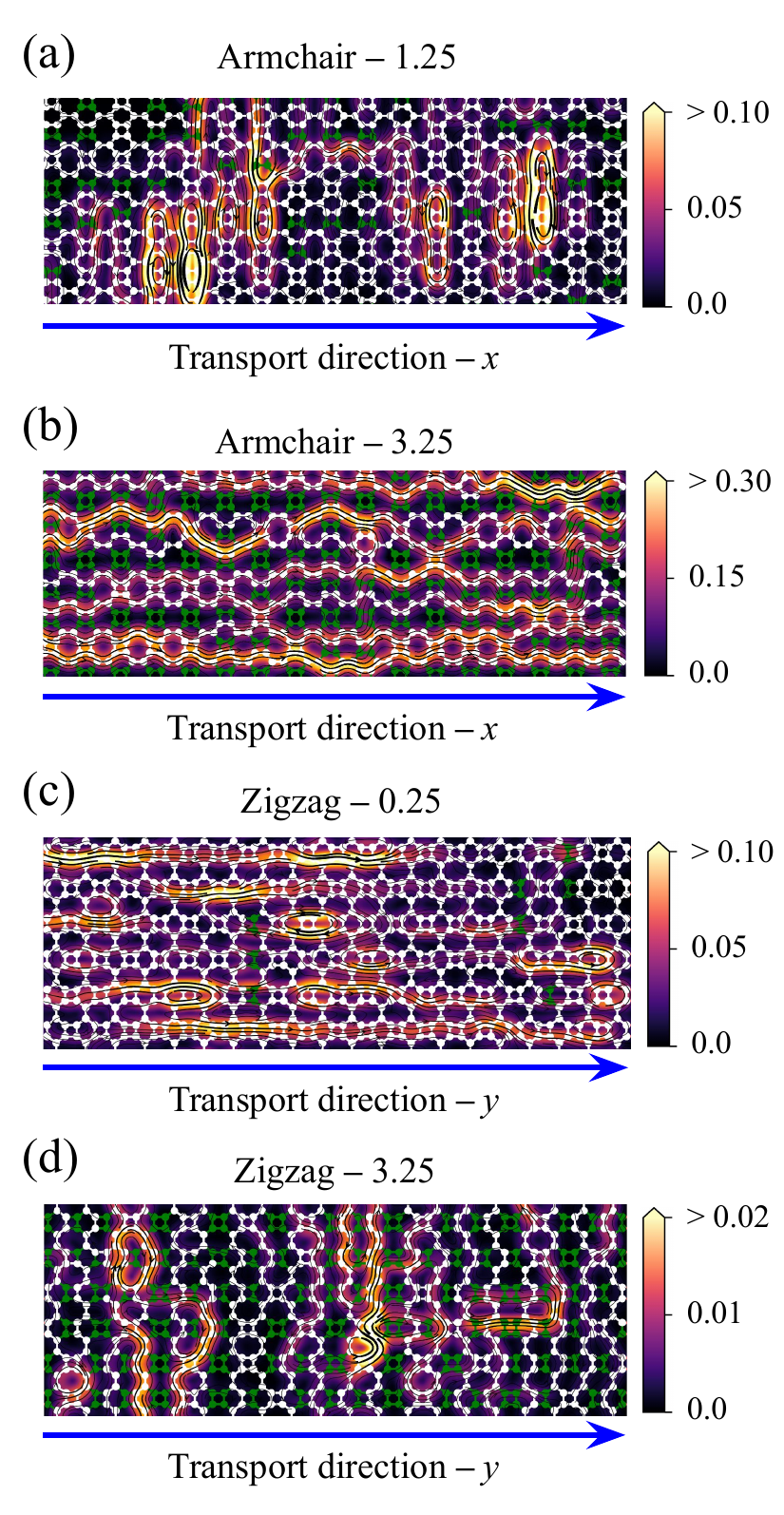}
    \caption{Distribution of the current density in disordered fluorinated  BPN, F$_{2.0}$/F$^\text{d}_{x_\text{ad}}$/F$_{2.0}$, along the armchair direction ($I_x$) for $x_\text{ad}$ of 1.25 (a) and 3.25 (b); along the zigzag direction ($I_y$) for $x_\text{ad}$ of 0.25 (c) and
 3.25 (d). Green circles indicate the fluorinated sites.}  
    \label{fig:current_density}
\end{figure}

A qualitative understanding of these findings can be obtained from the calculated spatial distribution of the current density along the armchair ($I_x$) and zigzag ($I_y$) directions In  Fig.~\ref{fig:current_density}, we present the current densities $I_x$ and $I_y$ in F$_{2.0}$/F$_{x_\text{ad}}$/F$_{2.0}$.

We begin by analyzing the current densities along the armchair ($\hat x$) and zigzag ($\hat y$) directions for $x_\text{ad}=1.25$ [minimum value of $I_x$, Fig.~\ref{fig:rand_currents}(b)] and $0.25$ [maximum value of $I_y$, Fig.~\ref{fig:rand_currents}(d)], respectively, as shown in Figs.~\ref{fig:current_density}(a) and (c). In the former case, we observe a predominance of current-density components perpendicular to the transport direction, whereas in Fig.~\ref{fig:current_density}(c) the current density is predominantly aligned parallel to the transport direction. These results are consistent with the directional dependence of electronic transport in pristine biphenylene [Fig.~\ref{fig:bpn}(c)], where $\mathcal{T}_y > \mathcal{T}_x$, as discussed in Ref.~\cite{PCCPkuritza2024}.

As the fluorine adatom concentration increases to $x_\text{ad}=3.25$ [maximum value of $I_x$, Fig.~\ref{fig:rand_currents}(b)], with $\eta<1$ [Fig.~\ref{fig:rand_currents}(f)], we observe the preferential formation of quasi-continuous fluorine-adatom lines along the armchair direction, i.e., parallel to $I_x$ [Fig.~\ref{fig:current_density}(b)]. We infer that the enhancement of the current for $x_\text{ad}=3.25$, shown in Figs.~\ref{fig:rand_currents}(a) and (b), arises from the formation of electronic transmission channels, mediated by the overlap of C-$\pi$ orbitals, induced by electronic confinement effects associated with these quasi-linear fluorine-adatom structures aligned parallel to the transport direction. In contrast, the preferential formation of such quasi-linear fluorine-adatom structures along the AC direction ($\hat x$) produces a blocking effect on the current density in $I_y$ [Fig.~\ref{fig:current_density}(d)], thereby leading to $I(V,x_{\mathrm{ad}})\rightarrow0$ [Figs.~\ref{fig:rand_currents}(c) and (d)]. Similar results were obtained in F$_{1.5}$/F$_{x_\text{ad}}$/F$_{1.5}$.

Since the spatial distribution of fluorine adatoms governs both the current anisotropy, $\eta=I_y/I_x$, and the nonmonotonic behavior of $I_x$, it is instructive to examine the energetic contributions underlying the distribution of fluorine adatoms and their implications for the electronic transport properties. The disordered distribution of fluorine adatoms is determined by the configurational energy, $E_\text{config}$, defined in Eq.~\eqref{B1}. The first term, $J_{ij}$, describes the correlation between pairs of fluorine adatoms, whereas the second term, $E_i^B$, corresponds to the interaction between the fluorine adatoms and the BPN surface.

\begin{figure}
    \centering
    \includegraphics[width=0.8\columnwidth]{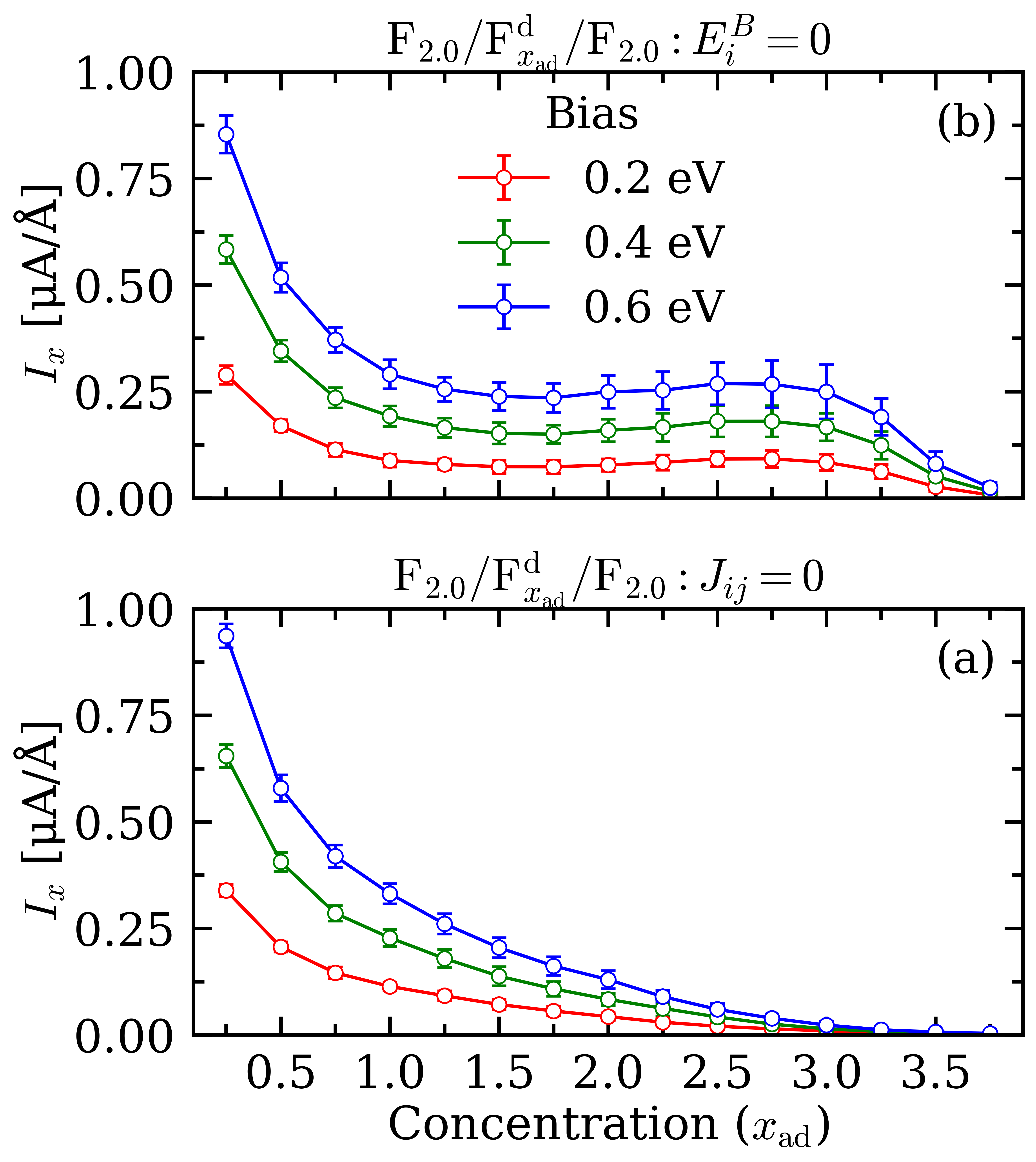}
    \caption{Current as a function of the concentration of fluorine adatoms ($x_\text{ad}$) and bias voltage in disordered fluorinated BPN along the armchair
    direction ($I_x$) connected to F$_{2.0}$/BPN leads for (a) correlated, $E^B_i=0$ and $J_{ij}\neq 0$, and (b) non-correlated, $E^B_i\neq 0$ and $J_{ij} = 0$, distribution of fluorine adatoms on the BPN scatterer.}
    \label{fig:uncorrelated}
\end{figure}

We examine the current in F$_{2.0}$/F$^\text{d}_{x_\text{ad}}$/F$_{2.0}$ along the armchair direction for two hypothetical fluorine-adatom distributions: (i) neglecting the binding-energy contribution, $E_i^B$, in $E_\text{config}$ while retaining the interaction energy between fluorine adatoms, $J_{ij}$, and (ii) neglecting $J_{ij}$ while retaining the binding energy between the fluorine adatoms and the biphenylene surface, $E_i^B$.
For $E_i^B = 0$ and $J_{ij} \neq 0$, a small but finite current persists around $x_\text{ad}=3.25$ [Fig.~\ref{fig:uncorrelated}(a)]. In contrast, when $J_{ij} = 0$ and $E_i^B \neq 0$, the current decays exponentially toward zero [Fig.~\ref{fig:uncorrelated}(b)].

These results indicate that correlations between fluorine adatoms, described by $J_{ij}$, provide the dominant contribution to the formation of quasi-linear fluorine-adatom arrays on biphenylene. Such correlated structures lead to the exponential suppression of the current along the zigzag direction, $I_y$, as well as to the nonmonotonic behavior of the armchair current, $I_x$, mediated by the emergence of armchair-oriented C-$\pi$ transport channels on the fluorine-covered biphenylene surface.

\section{Conclusion}

In summary, we investigate the electronic and transport properties of fluorinated biphenylene networks (F/BPN) consisting of F$_n$/BPN leads ($n=1.5$ and $2.0$) connected to {\it (1)} pristine BPN, {\it (2)} ordered fluorinated BPN, and {\it (3)} disordered fluorinated BPN. The calculations were performed by combining density functional theory, Wannier-based tight-binding Hamiltonians, and quantum transport simulations within the Landauer-B\"uttiker formalism. Our results demonstrate that fluorination provides an effective route for tailoring both the electronic structure and the transport response of BPN.

For pristine BPN, F$_n$/BPN/F$_n$, we identified the emergence of negative differential resistance (NDR) in the armchair current ($I_x$), whereas the zigzag current ($I_y$) exhibits a nearly Ohmic behavior. A comparison between $I_x$ and $I_y$ further reveals that the NDR induces an inversion of the current anisotropy factor, $\eta=I_y/I_x$, from $\eta<1$ to $\eta>1$. In contrast, ordered fluorination, F$_n$/F$^\text{o}_n$/F$_n$, leads to NDR in both transport directions for F$_{1.5}$/F$^\text{o}_{1.5}$/F$_{1.5}$. For F$_{2.0}$/F$^\text{o}_{2.0}$/F$_{2.0}$, however, no NDR is observed; instead, both $I_x$ and $I_y$ exhibit current saturation with increasing bias voltage, while the anisotropy factor remains below unity ($\eta<1$).

Introducing disorder into the fluorinated BPN suppresses the NDR effect, driving the system toward a nearly Ohmic transport regime. Despite this suppression, the transport anisotropy remains strongly dependent on both the fluorine concentration and the spatial distribution of the adatoms. Most importantly, at high fluorine concentrations, we identified an unconventional transport regime characterized by a nonmonotonic dependence of the armchair current, $I_x$, on adatom concentration. Based on our total-energy calculations, we attribute this behavior to the formation of correlated quasi-linear fluorine-adatom structures that promote armchair-oriented C-$\pi$ transport channels while simultaneously suppressing transport along the zigzag direction.

These findings demonstrate that correlated chemical disorder in fluorinated biphenylene does not merely degrade electronic transport but can instead generate emergent conducting pathways and unexpected anisotropic transport regimes. More broadly, they highlight the potential of controlled adatom correlations as a strategy for tailoring directional charge transport in two-dimensional carbon-based materials.

\begin{acknowledgments}
The authors acknowledge financial support from the Brazilian agencies FAPESP (grants 23/09820-2 and 24/00989-7), CNPq, INCT-Nanocarbono e Materiais 2D, INCT-Materials Informatics, and the LNCC (SDumont, Project SCAFMat2) for computer time.
\end{acknowledgments}

\appendix

\bibliography{biblio}

\begin{thebibliography}{42}%
\makeatletter
\providecommand \@ifxundefined [1]{%
 \@ifx{#1\undefined}
}%
\providecommand \@ifnum [1]{%
 \ifnum #1\expandafter \@firstoftwo
 \else \expandafter \@secondoftwo
 \fi
}%
\providecommand \@ifx [1]{%
 \ifx #1\expandafter \@firstoftwo
 \else \expandafter \@secondoftwo
 \fi
}%
\providecommand \natexlab [1]{#1}%
\providecommand \enquote  [1]{``#1''}%
\providecommand \bibnamefont  [1]{#1}%
\providecommand \bibfnamefont [1]{#1}%
\providecommand \citenamefont [1]{#1}%
\providecommand \href@noop [0]{\@secondoftwo}%
\providecommand \href [0]{\begingroup \@sanitize@url \@href}%
\providecommand \@href[1]{\@@startlink{#1}\@@href}%
\providecommand \@@href[1]{\endgroup#1\@@endlink}%
\providecommand \@sanitize@url [0]{\catcode `\\12\catcode `\$12\catcode `\&12\catcode `\#12\catcode `\^12\catcode `\_12\catcode `\%12\relax}%
\providecommand \@@startlink[1]{}%
\providecommand \@@endlink[0]{}%
\providecommand \url  [0]{\begingroup\@sanitize@url \@url }%
\providecommand \@url [1]{\endgroup\@href {#1}{\urlprefix }}%
\providecommand \urlprefix  [0]{URL }%
\providecommand \Eprint [0]{\href }%
\providecommand \doibase [0]{https://doi.org/}%
\providecommand \selectlanguage [0]{\@gobble}%
\providecommand \bibinfo  [0]{\@secondoftwo}%
\providecommand \bibfield  [0]{\@secondoftwo}%
\providecommand \translation [1]{[#1]}%
\providecommand \BibitemOpen [0]{}%
\providecommand \bibitemStop [0]{}%
\providecommand \bibitemNoStop [0]{.\EOS\space}%
\providecommand \EOS [0]{\spacefactor3000\relax}%
\providecommand \BibitemShut  [1]{\csname bibitem#1\endcsname}%
\let\auto@bib@innerbib\@empty
\bibitem [{\citenamefont {Paupitz}\ \emph {et~al.}(2026)\citenamefont {Paupitz}, \citenamefont {Fonseca}, \citenamefont {Bessa}, \citenamefont {Fabris}, \citenamefont {da~Cunha}, \citenamefont {Machado}, \citenamefont {Pereira}, \citenamefont {Ribeiro},\ and\ \citenamefont {Galvão}}]{paupitz2026concise}%
  \BibitemOpen
  \bibfield  {author} {\bibinfo {author} {\bibfnamefont {R.}~\bibnamefont {Paupitz}}, \bibinfo {author} {\bibfnamefont {A.}~\bibnamefont {Fonseca}}, \bibinfo {author} {\bibfnamefont {M.}~\bibnamefont {Bessa}}, \bibinfo {author} {\bibfnamefont {G.}~\bibnamefont {Fabris}}, \bibinfo {author} {\bibfnamefont {W.}~\bibnamefont {da~Cunha}}, \bibinfo {author} {\bibfnamefont {L.}~\bibnamefont {Machado}}, \bibinfo {author} {\bibfnamefont {M.}~\bibnamefont {Pereira}}, \bibinfo {author} {\bibfnamefont {L.}~\bibnamefont {Ribeiro}},\ and\ \bibinfo {author} {\bibfnamefont {D.}~\bibnamefont {Galvão}},\ }\bibfield  {title} {\bibinfo {title} {{A concise review of recently synthesized 2D carbon allotropes: Amorphous carbon, graphynes, biphenylene and fullerene networks}},\ }\href {https://doi.org/10.1016/j.carbon.2026.121320} {\bibfield  {journal} {\bibinfo  {journal} {Carbon}\ }\textbf {\bibinfo {volume} {252}},\ \bibinfo {pages} {121320} (\bibinfo {year} {2026})}\BibitemShut {NoStop}%
\bibitem [{\citenamefont {Ahmad}\ \emph {et~al.}(2024)\citenamefont {Ahmad}, \citenamefont {Mahmood},\ and\ \citenamefont {Muhmood}}]{ahmad2024carbon}%
  \BibitemOpen
  \bibfield  {author} {\bibinfo {author} {\bibfnamefont {F.}~\bibnamefont {Ahmad}}, \bibinfo {author} {\bibfnamefont {A.}~\bibnamefont {Mahmood}},\ and\ \bibinfo {author} {\bibfnamefont {T.}~\bibnamefont {Muhmood}},\ }\bibinfo {title} {Carbon allotropes: Basics, properties and applications},\ in\ \href {https://doi.org/10.1021/bk-2024-1491.ch001} {\emph {\bibinfo {booktitle} {Heteroatom-Doped Carbon Allotropes: Progress in Synthesis, Characterization, and Applications}}}\ (\bibinfo  {publisher} {American Chemical Society},\ \bibinfo {year} {2024})\ p.\ \bibinfo {pages} {1–18}\BibitemShut {NoStop}%
\bibitem [{\citenamefont {Liu}\ \emph {et~al.}(2012)\citenamefont {Liu}, \citenamefont {Wang}, \citenamefont {Huang}, \citenamefont {Guo},\ and\ \citenamefont {Chen}}]{liu2012structural}%
  \BibitemOpen
  \bibfield  {author} {\bibinfo {author} {\bibfnamefont {Y.}~\bibnamefont {Liu}}, \bibinfo {author} {\bibfnamefont {G.}~\bibnamefont {Wang}}, \bibinfo {author} {\bibfnamefont {Q.}~\bibnamefont {Huang}}, \bibinfo {author} {\bibfnamefont {L.}~\bibnamefont {Guo}},\ and\ \bibinfo {author} {\bibfnamefont {X.}~\bibnamefont {Chen}},\ }\bibfield  {title} {\bibinfo {title} {{Structural and Electronic Properties of $T$ Graphene: A Two-Dimensional Carbon Allotrope with Tetrarings}},\ }\href {https://doi.org/10.1103/PhysRevLett.108.225505} {\bibfield  {journal} {\bibinfo  {journal} {Phys. Rev. Lett.}\ }\textbf {\bibinfo {volume} {108}},\ \bibinfo {pages} {225505} (\bibinfo {year} {2012})}\BibitemShut {NoStop}%
\bibitem [{\citenamefont {Li}\ \emph {et~al.}(2023)\citenamefont {Li}, \citenamefont {Lim}, \citenamefont {Lv}, \citenamefont {Li}, \citenamefont {Kang},\ and\ \citenamefont {Lee}}]{li2023graphynes}%
  \BibitemOpen
  \bibfield  {author} {\bibinfo {author} {\bibfnamefont {H.}~\bibnamefont {Li}}, \bibinfo {author} {\bibfnamefont {J.~H.}\ \bibnamefont {Lim}}, \bibinfo {author} {\bibfnamefont {Y.}~\bibnamefont {Lv}}, \bibinfo {author} {\bibfnamefont {N.}~\bibnamefont {Li}}, \bibinfo {author} {\bibfnamefont {B.}~\bibnamefont {Kang}},\ and\ \bibinfo {author} {\bibfnamefont {J.~Y.}\ \bibnamefont {Lee}},\ }\bibfield  {title} {\bibinfo {title} {{Graphynes and Graphdiynes for Energy Storage and Catalytic Utilization: Theoretical Insights into Recent Advances}},\ }\href {https://doi.org/10.1021/acs.chemrev.2c00729} {\bibfield  {journal} {\bibinfo  {journal} {Chemical Reviews}\ }\textbf {\bibinfo {volume} {123}},\ \bibinfo {pages} {4795–4854} (\bibinfo {year} {2023})}\BibitemShut {NoStop}%
\bibitem [{\citenamefont {Liu}\ \emph {et~al.}(2024)\citenamefont {Liu}, \citenamefont {Wang}, \citenamefont {Yu},\ and\ \citenamefont {Wang}}]{liu2024graphyne}%
  \BibitemOpen
  \bibfield  {author} {\bibinfo {author} {\bibfnamefont {Q.}~\bibnamefont {Liu}}, \bibinfo {author} {\bibfnamefont {X.}~\bibnamefont {Wang}}, \bibinfo {author} {\bibfnamefont {J.}~\bibnamefont {Yu}},\ and\ \bibinfo {author} {\bibfnamefont {J.}~\bibnamefont {Wang}},\ }\bibfield  {title} {\bibinfo {title} {{Graphyne and graphdiyne nanoribbons: from their structures and properties to potential applications}},\ }\href {https://doi.org/10.1039/d3cp04393b} {\bibfield  {journal} {\bibinfo  {journal} {Physical Chemistry Chemical Physics}\ }\textbf {\bibinfo {volume} {26}},\ \bibinfo {pages} {1541–1563} (\bibinfo {year} {2024})}\BibitemShut {NoStop}%
\bibitem [{\citenamefont {Koizumi}\ \emph {et~al.}(2024)\citenamefont {Koizumi}, \citenamefont {Phan}, \citenamefont {Nishigomi},\ and\ \citenamefont {Wakabayashi}}]{koizumi2024topological}%
  \BibitemOpen
  \bibfield  {author} {\bibinfo {author} {\bibfnamefont {K.}~\bibnamefont {Koizumi}}, \bibinfo {author} {\bibfnamefont {H.~T.}\ \bibnamefont {Phan}}, \bibinfo {author} {\bibfnamefont {K.}~\bibnamefont {Nishigomi}},\ and\ \bibinfo {author} {\bibfnamefont {K.}~\bibnamefont {Wakabayashi}},\ }\bibfield  {title} {\bibinfo {title} {{Topological edge and corner states in the biphenylene network}},\ }\href {https://doi.org/10.1103/PhysRevB.109.035431} {\bibfield  {journal} {\bibinfo  {journal} {Phys. Rev. B}\ }\textbf {\bibinfo {volume} {109}},\ \bibinfo {pages} {035431} (\bibinfo {year} {2024})}\BibitemShut {NoStop}%
\bibitem [{\citenamefont {Son}\ \emph {et~al.}(2022)\citenamefont {Son}, \citenamefont {Jin},\ and\ \citenamefont {Kim}}]{son2022magnetic}%
  \BibitemOpen
  \bibfield  {author} {\bibinfo {author} {\bibfnamefont {Y.-W.}\ \bibnamefont {Son}}, \bibinfo {author} {\bibfnamefont {H.}~\bibnamefont {Jin}},\ and\ \bibinfo {author} {\bibfnamefont {S.}~\bibnamefont {Kim}},\ }\bibfield  {title} {\bibinfo {title} {{Magnetic Ordering, Anomalous Lifshitz Transition, and Topological Grain Boundaries in Two-Dimensional Biphenylene Network}},\ }\href {https://doi.org/10.1021/acs.nanolett.2c00528} {\bibfield  {journal} {\bibinfo  {journal} {Nano Letters}\ }\textbf {\bibinfo {volume} {22}},\ \bibinfo {pages} {3112–3117} (\bibinfo {year} {2022})}\BibitemShut {NoStop}%
\bibitem [{\citenamefont {Crasto~de Lima}\ \emph {et~al.}(2019)\citenamefont {Crasto~de Lima}, \citenamefont {Ferreira},\ and\ \citenamefont {Miwa}}]{PCCPcrasto2019}%
  \BibitemOpen
  \bibfield  {author} {\bibinfo {author} {\bibfnamefont {F.}~\bibnamefont {Crasto~de Lima}}, \bibinfo {author} {\bibfnamefont {G.~J.}\ \bibnamefont {Ferreira}},\ and\ \bibinfo {author} {\bibfnamefont {R.~H.}\ \bibnamefont {Miwa}},\ }\bibfield  {title} {\bibinfo {title} {{Topological flat band, Dirac fermions and quantum spin Hall phase in 2D Archimedean lattices}},\ }\href {https://doi.org/10.1039/c9cp04760c} {\bibfield  {journal} {\bibinfo  {journal} {Physical Chemistry Chemical Physics}\ }\textbf {\bibinfo {volume} {21}},\ \bibinfo {pages} {22344–22350} (\bibinfo {year} {2019})}\BibitemShut {NoStop}%
\bibitem [{\citenamefont {Fan}\ \emph {et~al.}(2021)\citenamefont {Fan}, \citenamefont {Yan}, \citenamefont {Tripp}, \citenamefont {Krejčí}, \citenamefont {Dimosthenous}, \citenamefont {Kachel}, \citenamefont {Chen}, \citenamefont {Foster}, \citenamefont {Koert}, \citenamefont {Liljeroth},\ and\ \citenamefont {Gottfried}}]{SCIENCEfan2021}%
  \BibitemOpen
  \bibfield  {author} {\bibinfo {author} {\bibfnamefont {Q.}~\bibnamefont {Fan}}, \bibinfo {author} {\bibfnamefont {L.}~\bibnamefont {Yan}}, \bibinfo {author} {\bibfnamefont {M.~W.}\ \bibnamefont {Tripp}}, \bibinfo {author} {\bibfnamefont {O.}~\bibnamefont {Krejčí}}, \bibinfo {author} {\bibfnamefont {S.}~\bibnamefont {Dimosthenous}}, \bibinfo {author} {\bibfnamefont {S.~R.}\ \bibnamefont {Kachel}}, \bibinfo {author} {\bibfnamefont {M.}~\bibnamefont {Chen}}, \bibinfo {author} {\bibfnamefont {A.~S.}\ \bibnamefont {Foster}}, \bibinfo {author} {\bibfnamefont {U.}~\bibnamefont {Koert}}, \bibinfo {author} {\bibfnamefont {P.}~\bibnamefont {Liljeroth}},\ and\ \bibinfo {author} {\bibfnamefont {J.~M.}\ \bibnamefont {Gottfried}},\ }\bibfield  {title} {\bibinfo {title} {Biphenylene network: A nonbenzenoid carbon allotrope},\ }\href {https://doi.org/10.1126/science.abg4509} {\bibfield  {journal} {\bibinfo  {journal} {Science}\ }\textbf {\bibinfo {volume} {372}},\ \bibinfo {pages} {852–856} (\bibinfo {year} {2021})}\BibitemShut {NoStop}%
\bibitem [{\citenamefont {Tong}\ \emph {et~al.}(2022)\citenamefont {Tong}, \citenamefont {Pecchia}, \citenamefont {Yam}, \citenamefont {Dumitrică},\ and\ \citenamefont {Frauenheim}}]{AEMtong2022}%
  \BibitemOpen
  \bibfield  {author} {\bibinfo {author} {\bibfnamefont {Z.}~\bibnamefont {Tong}}, \bibinfo {author} {\bibfnamefont {A.}~\bibnamefont {Pecchia}}, \bibinfo {author} {\bibfnamefont {C.}~\bibnamefont {Yam}}, \bibinfo {author} {\bibfnamefont {T.}~\bibnamefont {Dumitrică}},\ and\ \bibinfo {author} {\bibfnamefont {T.}~\bibnamefont {Frauenheim}},\ }\bibfield  {title} {\bibinfo {title} {{Ultrahigh Electron Thermal Conductivity in T‐Graphene, Biphenylene, and Net‐Graphene}},\ }\bibfield  {journal} {\bibinfo  {journal} {Advanced Energy Materials}\ }\textbf {\bibinfo {volume} {12}},\ \href {https://doi.org/10.1002/aenm.202200657} {10.1002/aenm.202200657} (\bibinfo {year} {2022})\BibitemShut {NoStop}%
\bibitem [{\citenamefont {Liu}\ \emph {et~al.}(2021{\natexlab{a}})\citenamefont {Liu}, \citenamefont {Li}, \citenamefont {Zhang}, \citenamefont {Tu}, \citenamefont {Zhang}, \citenamefont {Zhang}, \citenamefont {Wang},\ and\ \citenamefont {Singh}}]{PRBliu2021}%
  \BibitemOpen
  \bibfield  {author} {\bibinfo {author} {\bibfnamefont {P.-F.}\ \bibnamefont {Liu}}, \bibinfo {author} {\bibfnamefont {J.}~\bibnamefont {Li}}, \bibinfo {author} {\bibfnamefont {C.}~\bibnamefont {Zhang}}, \bibinfo {author} {\bibfnamefont {X.-H.}\ \bibnamefont {Tu}}, \bibinfo {author} {\bibfnamefont {J.}~\bibnamefont {Zhang}}, \bibinfo {author} {\bibfnamefont {P.}~\bibnamefont {Zhang}}, \bibinfo {author} {\bibfnamefont {B.-T.}\ \bibnamefont {Wang}},\ and\ \bibinfo {author} {\bibfnamefont {D.~J.}\ \bibnamefont {Singh}},\ }\bibfield  {title} {\bibinfo {title} {{Type-II Dirac cones and electron-phonon interaction in monolayer biphenylene from first-principles calculations}},\ }\href {https://doi.org/10.1103/PhysRevB.104.235422} {\bibfield  {journal} {\bibinfo  {journal} {Phys. Rev. B}\ }\textbf {\bibinfo {volume} {104}},\ \bibinfo {pages} {235422} (\bibinfo {year} {2021}{\natexlab{a}})}\BibitemShut {NoStop}%
\bibitem [{\citenamefont {Luo}\ \emph {et~al.}(2021)\citenamefont {Luo}, \citenamefont {Ren}, \citenamefont {Xu}, \citenamefont {Yu}, \citenamefont {Wang},\ and\ \citenamefont {Sun}}]{SRluo2021}%
  \BibitemOpen
  \bibfield  {author} {\bibinfo {author} {\bibfnamefont {Y.}~\bibnamefont {Luo}}, \bibinfo {author} {\bibfnamefont {C.}~\bibnamefont {Ren}}, \bibinfo {author} {\bibfnamefont {Y.}~\bibnamefont {Xu}}, \bibinfo {author} {\bibfnamefont {J.}~\bibnamefont {Yu}}, \bibinfo {author} {\bibfnamefont {S.}~\bibnamefont {Wang}},\ and\ \bibinfo {author} {\bibfnamefont {M.}~\bibnamefont {Sun}},\ }\bibfield  {title} {\bibinfo {title} {{A first principles investigation on the structural, mechanical, electronic, and catalytic properties of biphenylene}},\ }\bibfield  {journal} {\bibinfo  {journal} {Scientific Reports}\ }\textbf {\bibinfo {volume} {11}},\ \href {https://doi.org/10.1038/s41598-021-98261-9} {10.1038/s41598-021-98261-9} (\bibinfo {year} {2021})\BibitemShut {NoStop}%
\bibitem [{\citenamefont {Liu}\ \emph {et~al.}(2021{\natexlab{b}})\citenamefont {Liu}, \citenamefont {Jing},\ and\ \citenamefont {Li}}]{JPCLliu2021}%
  \BibitemOpen
  \bibfield  {author} {\bibinfo {author} {\bibfnamefont {T.}~\bibnamefont {Liu}}, \bibinfo {author} {\bibfnamefont {Y.}~\bibnamefont {Jing}},\ and\ \bibinfo {author} {\bibfnamefont {Y.}~\bibnamefont {Li}},\ }\bibfield  {title} {\bibinfo {title} {{Two-Dimensional Biphenylene: A Graphene Allotrope with Superior Activity toward Electrochemical Oxygen Reduction Reaction}},\ }\href {https://doi.org/10.1021/acs.jpclett.1c03851} {\bibfield  {journal} {\bibinfo  {journal} {The Journal of Physical Chemistry Letters}\ }\textbf {\bibinfo {volume} {12}},\ \bibinfo {pages} {12230–12234} (\bibinfo {year} {2021}{\natexlab{b}})}\BibitemShut {NoStop}%
\bibitem [{\citenamefont {Lebre}\ \emph {et~al.}(2025)\citenamefont {Lebre}, \citenamefont {Pacine}, \citenamefont {Lima}, \citenamefont {Cotta},\ and\ \citenamefont {de~Oliveira}}]{ACSANMlebre2025}%
  \BibitemOpen
  \bibfield  {author} {\bibinfo {author} {\bibfnamefont {M.~P.}\ \bibnamefont {Lebre}}, \bibinfo {author} {\bibfnamefont {D.}~\bibnamefont {Pacine}}, \bibinfo {author} {\bibfnamefont {E.~N.}\ \bibnamefont {Lima}}, \bibinfo {author} {\bibfnamefont {A.~A.~C.}\ \bibnamefont {Cotta}},\ and\ \bibinfo {author} {\bibfnamefont {I.~S.~S.}\ \bibnamefont {de~Oliveira}},\ }\bibfield  {title} {\bibinfo {title} {{First-Principles Study of Metal–Biphenylene Nanoscale Interfaces: Structural, Electronic, and Catalytic Properties}},\ }\bibfield  {journal} {\bibinfo  {journal} {ACS Applied Nano Materials}\ }\href {https://doi.org/10.1021/acsanm.5c02590} {10.1021/acsanm.5c02590} (\bibinfo {year} {2025})\BibitemShut {NoStop}%
\bibitem [{\citenamefont {Han}\ \emph {et~al.}(2022)\citenamefont {Han}, \citenamefont {Liu}, \citenamefont {Lv},\ and\ \citenamefont {Li}}]{PCCPhan2022}%
  \BibitemOpen
  \bibfield  {author} {\bibinfo {author} {\bibfnamefont {T.}~\bibnamefont {Han}}, \bibinfo {author} {\bibfnamefont {Y.}~\bibnamefont {Liu}}, \bibinfo {author} {\bibfnamefont {X.}~\bibnamefont {Lv}},\ and\ \bibinfo {author} {\bibfnamefont {F.}~\bibnamefont {Li}},\ }\bibfield  {title} {\bibinfo {title} {{Biphenylene monolayer: a novel nonbenzenoid carbon allotrope with potential application as an anode material for high-performance sodium-ion batteries}},\ }\href {https://doi.org/10.1039/d2cp00798c} {\bibfield  {journal} {\bibinfo  {journal} {Physical Chemistry Chemical Physics}\ }\textbf {\bibinfo {volume} {24}},\ \bibinfo {pages} {10712–10716} (\bibinfo {year} {2022})}\BibitemShut {NoStop}%
\bibitem [{\citenamefont {Kuritza}\ \emph {et~al.}(2024)\citenamefont {Kuritza}, \citenamefont {Miwa},\ and\ \citenamefont {Padilha}}]{PCCPkuritza2024}%
  \BibitemOpen
  \bibfield  {author} {\bibinfo {author} {\bibfnamefont {D.~P.}\ \bibnamefont {Kuritza}}, \bibinfo {author} {\bibfnamefont {R.~H.}\ \bibnamefont {Miwa}},\ and\ \bibinfo {author} {\bibfnamefont {J.~E.}\ \bibnamefont {Padilha}},\ }\bibfield  {title} {\bibinfo {title} {{Directional dependence of the electronic and transport properties of biphenylene under strain conditions}},\ }\href {https://doi.org/10.1039/d4cp00033a} {\bibfield  {journal} {\bibinfo  {journal} {Physical Chemistry Chemical Physics}\ }\textbf {\bibinfo {volume} {26}},\ \bibinfo {pages} {12142–12149} (\bibinfo {year} {2024})}\BibitemShut {NoStop}%
\bibitem [{\citenamefont {Feng}\ \emph {et~al.}(2016)\citenamefont {Feng}, \citenamefont {Long}, \citenamefont {Feng},\ and\ \citenamefont {Li}}]{ASfeng2016}%
  \BibitemOpen
  \bibfield  {author} {\bibinfo {author} {\bibfnamefont {W.}~\bibnamefont {Feng}}, \bibinfo {author} {\bibfnamefont {P.}~\bibnamefont {Long}}, \bibinfo {author} {\bibfnamefont {Y.}~\bibnamefont {Feng}},\ and\ \bibinfo {author} {\bibfnamefont {Y.}~\bibnamefont {Li}},\ }\bibfield  {title} {\bibinfo {title} {{Two‐Dimensional Fluorinated Graphene: Synthesis, Structures, Properties and Applications}},\ }\bibfield  {journal} {\bibinfo  {journal} {Advanced Science}\ }\textbf {\bibinfo {volume} {3}},\ \href {https://doi.org/10.1002/advs.201500413} {10.1002/advs.201500413} (\bibinfo {year} {2016})\BibitemShut {NoStop}%
\bibitem [{\citenamefont {Mo}\ \emph {et~al.}(2024)\citenamefont {Mo}, \citenamefont {Seo}, \citenamefont {Son}, \citenamefont {Kim}, \citenamefont {Rhim},\ and\ \citenamefont {Lee}}]{NLmo2024}%
  \BibitemOpen
  \bibfield  {author} {\bibinfo {author} {\bibfnamefont {S.}~\bibnamefont {Mo}}, \bibinfo {author} {\bibfnamefont {J.}~\bibnamefont {Seo}}, \bibinfo {author} {\bibfnamefont {S.-K.}\ \bibnamefont {Son}}, \bibinfo {author} {\bibfnamefont {S.}~\bibnamefont {Kim}}, \bibinfo {author} {\bibfnamefont {J.-W.}\ \bibnamefont {Rhim}},\ and\ \bibinfo {author} {\bibfnamefont {H.}~\bibnamefont {Lee}},\ }\bibfield  {title} {\bibinfo {title} {{Engineering Two-Dimensional Nodal Semimetals in Functionalized Biphenylene by Fluorine Adatoms}},\ }\href {https://doi.org/10.1021/acs.nanolett.4c00314} {\bibfield  {journal} {\bibinfo  {journal} {Nano Letters}\ }\textbf {\bibinfo {volume} {24}},\ \bibinfo {pages} {4885} (\bibinfo {year} {2024})}\BibitemShut {NoStop}%
\bibitem [{\citenamefont {Simonov}\ and\ \citenamefont {Goodwin}(2020)}]{NRCsimonov2020}%
  \BibitemOpen
  \bibfield  {author} {\bibinfo {author} {\bibfnamefont {A.}~\bibnamefont {Simonov}}\ and\ \bibinfo {author} {\bibfnamefont {A.~L.}\ \bibnamefont {Goodwin}},\ }\bibfield  {title} {\bibinfo {title} {{Designing disorder into crystalline materials}},\ }\href {https://doi.org/10.1038/s41570-020-00228-3} {\bibfield  {journal} {\bibinfo  {journal} {Nature Reviews Chemistry}\ }\textbf {\bibinfo {volume} {4}},\ \bibinfo {pages} {657–673} (\bibinfo {year} {2020})}\BibitemShut {NoStop}%
\bibitem [{\citenamefont {Li}\ \emph {et~al.}(2011)\citenamefont {Li}, \citenamefont {Hwang}, \citenamefont {Rossi},\ and\ \citenamefont {Das~Sarma}}]{PRLdassarma2011}%
  \BibitemOpen
  \bibfield  {author} {\bibinfo {author} {\bibfnamefont {Q.}~\bibnamefont {Li}}, \bibinfo {author} {\bibfnamefont {E.~H.}\ \bibnamefont {Hwang}}, \bibinfo {author} {\bibfnamefont {E.}~\bibnamefont {Rossi}},\ and\ \bibinfo {author} {\bibfnamefont {S.}~\bibnamefont {Das~Sarma}},\ }\bibfield  {title} {\bibinfo {title} {{Theory of 2D Transport in Graphene for Correlated Disorder}},\ }\href {https://doi.org/10.1103/PhysRevLett.107.156601} {\bibfield  {journal} {\bibinfo  {journal} {Phys. Rev. Lett.}\ }\textbf {\bibinfo {volume} {107}},\ \bibinfo {pages} {156601} (\bibinfo {year} {2011})}\BibitemShut {NoStop}%
\bibitem [{\citenamefont {Cho}\ \emph {et~al.}(2018)\citenamefont {Cho}, \citenamefont {Kończykowski}, \citenamefont {Teknowijoyo}, \citenamefont {Tanatar}, \citenamefont {Guss}, \citenamefont {Gartin}, \citenamefont {Wilde}, \citenamefont {Kreyssig}, \citenamefont {McQueeney}, \citenamefont {Goldman}, \citenamefont {Mishra}, \citenamefont {Hirschfeld},\ and\ \citenamefont {Prozorov}}]{NATCOMMcho2018}%
  \BibitemOpen
  \bibfield  {author} {\bibinfo {author} {\bibfnamefont {K.}~\bibnamefont {Cho}}, \bibinfo {author} {\bibfnamefont {M.}~\bibnamefont {Kończykowski}}, \bibinfo {author} {\bibfnamefont {S.}~\bibnamefont {Teknowijoyo}}, \bibinfo {author} {\bibfnamefont {M.~A.}\ \bibnamefont {Tanatar}}, \bibinfo {author} {\bibfnamefont {J.}~\bibnamefont {Guss}}, \bibinfo {author} {\bibfnamefont {P.~B.}\ \bibnamefont {Gartin}}, \bibinfo {author} {\bibfnamefont {J.~M.}\ \bibnamefont {Wilde}}, \bibinfo {author} {\bibfnamefont {A.}~\bibnamefont {Kreyssig}}, \bibinfo {author} {\bibfnamefont {R.~J.}\ \bibnamefont {McQueeney}}, \bibinfo {author} {\bibfnamefont {A.~I.}\ \bibnamefont {Goldman}}, \bibinfo {author} {\bibfnamefont {V.}~\bibnamefont {Mishra}}, \bibinfo {author} {\bibfnamefont {P.~J.}\ \bibnamefont {Hirschfeld}},\ and\ \bibinfo {author} {\bibfnamefont {R.}~\bibnamefont {Prozorov}},\ }\bibfield  {title} {\bibinfo {title} {{Using controlled disorder to probe the interplay between charge order and superconductivity in NbSe2}},\ }\bibfield  {journal} {\bibinfo  {journal} {Nature Communications}\ }\textbf {\bibinfo {volume} {9}},\ \href {https://doi.org/10.1038/s41467-018-05153-0} {10.1038/s41467-018-05153-0} (\bibinfo {year} {2018})\BibitemShut {NoStop}%
\bibitem [{\citenamefont {Neverov}\ \emph {et~al.}(2022)\citenamefont {Neverov}, \citenamefont {Lukyanov}, \citenamefont {Krasavin}, \citenamefont {Vagov},\ and\ \citenamefont {Croitoru}}]{COMMPHYSneverov2022}%
  \BibitemOpen
  \bibfield  {author} {\bibinfo {author} {\bibfnamefont {V.~D.}\ \bibnamefont {Neverov}}, \bibinfo {author} {\bibfnamefont {A.~E.}\ \bibnamefont {Lukyanov}}, \bibinfo {author} {\bibfnamefont {A.~V.}\ \bibnamefont {Krasavin}}, \bibinfo {author} {\bibfnamefont {A.}~\bibnamefont {Vagov}},\ and\ \bibinfo {author} {\bibfnamefont {M.~D.}\ \bibnamefont {Croitoru}},\ }\bibfield  {title} {\bibinfo {title} {{Correlated disorder as a way towards robust superconductivity}},\ }\bibfield  {journal} {\bibinfo  {journal} {Communications Physics}\ }\textbf {\bibinfo {volume} {5}},\ \href {https://doi.org/10.1038/s42005-022-00933-z} {10.1038/s42005-022-00933-z} (\bibinfo {year} {2022})\BibitemShut {NoStop}%
\bibitem [{\citenamefont {das Neves}\ \emph {et~al.}(2025)\citenamefont {das Neves}, \citenamefont {Bar\^ea}, \citenamefont {Perfecto}, \citenamefont {Bettini}, \citenamefont {de~Lima}, \citenamefont {Oliveira}, \citenamefont {Fazzio}, \citenamefont {Leite},\ and\ \citenamefont {Santhiago}}]{AMTdasneves2025}%
  \BibitemOpen
  \bibfield  {author} {\bibinfo {author} {\bibfnamefont {M.~F.~F.}\ \bibnamefont {das Neves}}, \bibinfo {author} {\bibfnamefont {H.~M.}\ \bibnamefont {Bar\^ea}}, \bibinfo {author} {\bibfnamefont {T.}~\bibnamefont {Perfecto}}, \bibinfo {author} {\bibfnamefont {J.}~\bibnamefont {Bettini}}, \bibinfo {author} {\bibfnamefont {F.~C.}\ \bibnamefont {de~Lima}}, \bibinfo {author} {\bibfnamefont {R.~F.}\ \bibnamefont {Oliveira}}, \bibinfo {author} {\bibfnamefont {A.}~\bibnamefont {Fazzio}}, \bibinfo {author} {\bibfnamefont {E.~R.}\ \bibnamefont {Leite}},\ and\ \bibinfo {author} {\bibfnamefont {M.}~\bibnamefont {Santhiago}},\ }\bibfield  {title} {\bibinfo {title} {{Room‐Temperature Tuning of Electrical Conductivity in Single MoS$_2$ Flakes via Nanoscale Amorphization by Focused Ion Beam}},\ }\bibfield  {journal} {\bibinfo  {journal} {Advanced Materials Technologies}\ }\textbf {\bibinfo {volume} {10}},\ \href {https://doi.org/10.1002/admt.202501505} {10.1002/admt.202501505} (\bibinfo {year} {2025})\BibitemShut {NoStop}%
\bibitem [{\citenamefont {Anderson}(1958)}]{PRanderson1958}%
  \BibitemOpen
  \bibfield  {author} {\bibinfo {author} {\bibfnamefont {P.~W.}\ \bibnamefont {Anderson}},\ }\bibfield  {title} {\bibinfo {title} {{Absence of Diffusion in Certain Random Lattices}},\ }\href {https://doi.org/10.1103/PhysRev.109.1492} {\bibfield  {journal} {\bibinfo  {journal} {Phys. Rev.}\ }\textbf {\bibinfo {volume} {109}},\ \bibinfo {pages} {1492} (\bibinfo {year} {1958})}\BibitemShut {NoStop}%
\bibitem [{\citenamefont {Izrailev}\ and\ \citenamefont {Krokhin}(1999)}]{PRLizrailev1999}%
  \BibitemOpen
  \bibfield  {author} {\bibinfo {author} {\bibfnamefont {F.~M.}\ \bibnamefont {Izrailev}}\ and\ \bibinfo {author} {\bibfnamefont {A.~A.}\ \bibnamefont {Krokhin}},\ }\bibfield  {title} {\bibinfo {title} {{Localization and the Mobility Edge in One-Dimensional Potentials with Correlated Disorder}},\ }\href {https://doi.org/10.1103/PhysRevLett.82.4062} {\bibfield  {journal} {\bibinfo  {journal} {Phys. Rev. Lett.}\ }\textbf {\bibinfo {volume} {82}},\ \bibinfo {pages} {4062} (\bibinfo {year} {1999})}\BibitemShut {NoStop}%
\bibitem [{\citenamefont {Bodyfelt}\ \emph {et~al.}(2014)\citenamefont {Bodyfelt}, \citenamefont {Leykam}, \citenamefont {Danieli}, \citenamefont {Yu},\ and\ \citenamefont {Flach}}]{PRLbodyfelt2014}%
  \BibitemOpen
  \bibfield  {author} {\bibinfo {author} {\bibfnamefont {J.~D.}\ \bibnamefont {Bodyfelt}}, \bibinfo {author} {\bibfnamefont {D.}~\bibnamefont {Leykam}}, \bibinfo {author} {\bibfnamefont {C.}~\bibnamefont {Danieli}}, \bibinfo {author} {\bibfnamefont {X.}~\bibnamefont {Yu}},\ and\ \bibinfo {author} {\bibfnamefont {S.}~\bibnamefont {Flach}},\ }\bibfield  {title} {\bibinfo {title} {{Flatbands under Correlated Perturbations}},\ }\href {https://doi.org/10.1103/PhysRevLett.113.236403} {\bibfield  {journal} {\bibinfo  {journal} {Phys. Rev. Lett.}\ }\textbf {\bibinfo {volume} {113}},\ \bibinfo {pages} {236403} (\bibinfo {year} {2014})}\BibitemShut {NoStop}%
\bibitem [{\citenamefont {Toloza~Sandoval}\ \emph {et~al.}(2025)\citenamefont {Toloza~Sandoval}, \citenamefont {Araújo}, \citenamefont {Crasto~de Lima},\ and\ \citenamefont {Fazzio}}]{PEtolozaSandoval2025}%
  \BibitemOpen
  \bibfield  {author} {\bibinfo {author} {\bibfnamefont {M.}~\bibnamefont {Toloza~Sandoval}}, \bibinfo {author} {\bibfnamefont {A.}~\bibnamefont {Araújo}}, \bibinfo {author} {\bibfnamefont {F.}~\bibnamefont {Crasto~de Lima}},\ and\ \bibinfo {author} {\bibfnamefont {A.}~\bibnamefont {Fazzio}},\ }\bibfield  {title} {\bibinfo {title} {{Transport fingerprints of helical edge states in Sierpiński tapestries}},\ }\href {https://doi.org/10.1016/j.physe.2024.116097} {\bibfield  {journal} {\bibinfo  {journal} {Physica E: Low-dimensional Systems and Nanostructures}\ }\textbf {\bibinfo {volume} {165}},\ \bibinfo {pages} {116097} (\bibinfo {year} {2025})}\BibitemShut {NoStop}%
\bibitem [{\citenamefont {Kresse}\ and\ \citenamefont {Furthm\"uller}(1996)}]{PRBkresse1996}%
  \BibitemOpen
  \bibfield  {author} {\bibinfo {author} {\bibfnamefont {G.}~\bibnamefont {Kresse}}\ and\ \bibinfo {author} {\bibfnamefont {J.}~\bibnamefont {Furthm\"uller}},\ }\bibfield  {title} {\bibinfo {title} {{Efficient iterative schemes for ab initio total-energy calculations using a plane-wave basis set}},\ }\href {https://doi.org/10.1103/PhysRevB.54.11169} {\bibfield  {journal} {\bibinfo  {journal} {Phys. Rev. B}\ }\textbf {\bibinfo {volume} {54}},\ \bibinfo {pages} {11169} (\bibinfo {year} {1996})}\BibitemShut {NoStop}%
\bibitem [{\citenamefont {Perdew}\ \emph {et~al.}(1996)\citenamefont {Perdew}, \citenamefont {Burke},\ and\ \citenamefont {Ernzerhof}}]{PRBperdew1996}%
  \BibitemOpen
  \bibfield  {author} {\bibinfo {author} {\bibfnamefont {J.~P.}\ \bibnamefont {Perdew}}, \bibinfo {author} {\bibfnamefont {K.}~\bibnamefont {Burke}},\ and\ \bibinfo {author} {\bibfnamefont {M.}~\bibnamefont {Ernzerhof}},\ }\bibfield  {title} {\bibinfo {title} {{Generalized Gradient Approximation Made Simple}},\ }\href {https://doi.org/10.1103/PhysRevLett.77.3865} {\bibfield  {journal} {\bibinfo  {journal} {Phys. Rev. Lett.}\ }\textbf {\bibinfo {volume} {77}},\ \bibinfo {pages} {3865} (\bibinfo {year} {1996})}\BibitemShut {NoStop}%
\bibitem [{\citenamefont {Grimme}\ \emph {et~al.}(2010)\citenamefont {Grimme}, \citenamefont {Antony}, \citenamefont {Ehrlich},\ and\ \citenamefont {Krieg}}]{JCPgrimme2010}%
  \BibitemOpen
  \bibfield  {author} {\bibinfo {author} {\bibfnamefont {S.}~\bibnamefont {Grimme}}, \bibinfo {author} {\bibfnamefont {J.}~\bibnamefont {Antony}}, \bibinfo {author} {\bibfnamefont {S.}~\bibnamefont {Ehrlich}},\ and\ \bibinfo {author} {\bibfnamefont {H.}~\bibnamefont {Krieg}},\ }\bibfield  {title} {\bibinfo {title} {{A consistent and accurate \textit{ab initio} parametrization of density functional dispersion correction (DFT-D) for the 94 elements H-Pu}},\ }\bibfield  {journal} {\bibinfo  {journal} {The Journal of Chemical Physics}\ }\textbf {\bibinfo {volume} {132}},\ \href {https://doi.org/10.1063/1.3382344} {10.1063/1.3382344} (\bibinfo {year} {2010})\BibitemShut {NoStop}%
\bibitem [{\citenamefont {Bl\"ochl}(1994)}]{PRBblochl1994}%
  \BibitemOpen
  \bibfield  {author} {\bibinfo {author} {\bibfnamefont {P.~E.}\ \bibnamefont {Bl\"ochl}},\ }\bibfield  {title} {\bibinfo {title} {{Projector augmented-wave method}},\ }\href {https://doi.org/10.1103/PhysRevB.50.17953} {\bibfield  {journal} {\bibinfo  {journal} {Phys. Rev. B}\ }\textbf {\bibinfo {volume} {50}},\ \bibinfo {pages} {17953} (\bibinfo {year} {1994})}\BibitemShut {NoStop}%
\bibitem [{\citenamefont {Kresse}\ and\ \citenamefont {Joubert}(1999)}]{PRBkresse1999}%
  \BibitemOpen
  \bibfield  {author} {\bibinfo {author} {\bibfnamefont {G.}~\bibnamefont {Kresse}}\ and\ \bibinfo {author} {\bibfnamefont {D.}~\bibnamefont {Joubert}},\ }\bibfield  {title} {\bibinfo {title} {{From ultrasoft pseudopotentials to the projector augmented-wave method}},\ }\href {https://doi.org/10.1103/PhysRevB.59.1758} {\bibfield  {journal} {\bibinfo  {journal} {Phys. Rev. B}\ }\textbf {\bibinfo {volume} {59}},\ \bibinfo {pages} {1758} (\bibinfo {year} {1999})}\BibitemShut {NoStop}%
\bibitem [{\citenamefont {Pizzi}\ \emph {et~al.}(2020)\citenamefont {Pizzi}, \citenamefont {Vitale}, \citenamefont {Arita}, \citenamefont {Blügel}, \citenamefont {Freimuth}, \citenamefont {G{\'{e}}ranton}, \citenamefont {Gibertini}, \citenamefont {Gresch}, \citenamefont {Johnson}, \citenamefont {Koretsune}, \citenamefont {Iba{\~{n}}ez-Azpiroz}, \citenamefont {Lee}, \citenamefont {Lihm}, \citenamefont {Marchand}, \citenamefont {Marrazzo}, \citenamefont {Mokrousov}, \citenamefont {Mustafa}, \citenamefont {Nohara}, \citenamefont {Nomura}, \citenamefont {Paulatto}, \citenamefont {Ponc{\'{e}}}, \citenamefont {Ponweiser}, \citenamefont {Qiao}, \citenamefont {Thöle}, \citenamefont {Tsirkin}, \citenamefont {Wierzbowska}, \citenamefont {Marzari}, \citenamefont {Vanderbilt}, \citenamefont {Souza}, \citenamefont {Mostofi},\ and\ \citenamefont {Yates}}]{JPCMpizzi2020}%
  \BibitemOpen
  \bibfield  {author} {\bibinfo {author} {\bibfnamefont {G.}~\bibnamefont {Pizzi}}, \bibinfo {author} {\bibfnamefont {V.}~\bibnamefont {Vitale}}, \bibinfo {author} {\bibfnamefont {R.}~\bibnamefont {Arita}}, \bibinfo {author} {\bibfnamefont {S.}~\bibnamefont {Blügel}}, \bibinfo {author} {\bibfnamefont {F.}~\bibnamefont {Freimuth}}, \bibinfo {author} {\bibfnamefont {G.}~\bibnamefont {G{\'{e}}ranton}}, \bibinfo {author} {\bibfnamefont {M.}~\bibnamefont {Gibertini}}, \bibinfo {author} {\bibfnamefont {D.}~\bibnamefont {Gresch}}, \bibinfo {author} {\bibfnamefont {C.}~\bibnamefont {Johnson}}, \bibinfo {author} {\bibfnamefont {T.}~\bibnamefont {Koretsune}}, \bibinfo {author} {\bibfnamefont {J.}~\bibnamefont {Iba{\~{n}}ez-Azpiroz}}, \bibinfo {author} {\bibfnamefont {H.}~\bibnamefont {Lee}}, \bibinfo {author} {\bibfnamefont {J.-M.}\ \bibnamefont {Lihm}}, \bibinfo {author} {\bibfnamefont {D.}~\bibnamefont {Marchand}}, \bibinfo {author} {\bibfnamefont {A.}~\bibnamefont {Marrazzo}}, \bibinfo {author} {\bibfnamefont {Y.}~\bibnamefont {Mokrousov}}, \bibinfo {author} {\bibfnamefont {J.~I.}\ \bibnamefont {Mustafa}}, \bibinfo {author} {\bibfnamefont {Y.}~\bibnamefont {Nohara}}, \bibinfo {author} {\bibfnamefont {Y.}~\bibnamefont {Nomura}}, \bibinfo {author} {\bibfnamefont {L.}~\bibnamefont {Paulatto}}, \bibinfo {author} {\bibfnamefont {S.}~\bibnamefont {Ponc{\'{e}}}}, \bibinfo {author} {\bibfnamefont {T.}~\bibnamefont {Ponweiser}}, \bibinfo {author} {\bibfnamefont {J.}~\bibnamefont {Qiao}}, \bibinfo {author} {\bibfnamefont {F.}~\bibnamefont {Thöle}}, \bibinfo {author} {\bibfnamefont {S.~S.}\ \bibnamefont {Tsirkin}}, \bibinfo {author} {\bibfnamefont {M.}~\bibnamefont {Wierzbowska}}, \bibinfo {author} {\bibfnamefont {N.}~\bibnamefont {Marzari}}, \bibinfo {author} {\bibfnamefont {D.}~\bibnamefont {Vanderbilt}}, \bibinfo {author} {\bibfnamefont {I.}~\bibnamefont {Souza}}, \bibinfo {author} {\bibfnamefont {A.~A.}\ \bibnamefont {Mostofi}},\ and\ \bibinfo {author} {\bibfnamefont {J.~R.}\ \bibnamefont {Yates}},\ }\bibfield  {title} {\bibinfo {title} {{Wannier90 as a community code: new features and applications}},\ }\href {https://doi.org/10.1088/1361-648x/ab51ff} {\bibfield  {journal} {\bibinfo  {journal} {Journal of Physics: Condensed Matter}\ }\textbf {\bibinfo {volume} {32}},\ \bibinfo {pages} {165902} (\bibinfo {year} {2020})}\BibitemShut {NoStop}%
\bibitem [{\citenamefont {Keldysh}(2023)}]{keldysh2024diagram}%
  \BibitemOpen
  \bibfield  {author} {\bibinfo {author} {\bibfnamefont {L.~V.}\ \bibnamefont {Keldysh}},\ }\bibinfo {title} {Diagram technique for nonequilibrium processes},\ in\ \href {https://doi.org/10.1142/9789811279461_0007} {\emph {\bibinfo {booktitle} {Selected Papers of Leonid V Keldysh}}}\ (\bibinfo  {publisher} {WORLD SCIENTIFIC},\ \bibinfo {year} {2023})\ p.\ \bibinfo {pages} {47–55}\BibitemShut {NoStop}%
\bibitem [{\citenamefont {Haug}\ and\ \citenamefont {Jauho}(2008)}]{haug2008quantum}%
  \BibitemOpen
  \bibfield  {author} {\bibinfo {author} {\bibfnamefont {H.}~\bibnamefont {Haug}}\ and\ \bibinfo {author} {\bibfnamefont {A.-P.}\ \bibnamefont {Jauho}},\ }\href {https://doi.org/10.1007/978-3-540-73564-9} {\emph {\bibinfo {title} {Quantum kinetics in transport and optics of semiconductors}}}\ (\bibinfo  {publisher} {Springer Berlin Heidelberg},\ \bibinfo {year} {2008})\BibitemShut {NoStop}%
\bibitem [{\citenamefont {Brandbyge}\ \emph {et~al.}(2002)\citenamefont {Brandbyge}, \citenamefont {Mozos}, \citenamefont {Ordej\'on}, \citenamefont {Taylor},\ and\ \citenamefont {Stokbro}}]{PRBbrandbyge2002}%
  \BibitemOpen
  \bibfield  {author} {\bibinfo {author} {\bibfnamefont {M.}~\bibnamefont {Brandbyge}}, \bibinfo {author} {\bibfnamefont {J.-L.}\ \bibnamefont {Mozos}}, \bibinfo {author} {\bibfnamefont {P.}~\bibnamefont {Ordej\'on}}, \bibinfo {author} {\bibfnamefont {J.}~\bibnamefont {Taylor}},\ and\ \bibinfo {author} {\bibfnamefont {K.}~\bibnamefont {Stokbro}},\ }\bibfield  {title} {\bibinfo {title} {{Density-functional method for nonequilibrium electron transport}},\ }\href {https://doi.org/10.1103/PhysRevB.65.165401} {\bibfield  {journal} {\bibinfo  {journal} {Phys. Rev. B}\ }\textbf {\bibinfo {volume} {65}},\ \bibinfo {pages} {165401} (\bibinfo {year} {2002})}\BibitemShut {NoStop}%
\bibitem [{Note1()}]{Note1}%
  \BibitemOpen
  \bibinfo {note} {Our TB approach does not solve the Keldysh nor the Poisson equations self-consistently, consequently, neglecting non-equilibrium charge redistribution. In order to get some insight about this configuration, we employ a linear potential drop in the scattering region matching the bias values at the electrodes}\BibitemShut {NoStop}%
\bibitem [{\citenamefont {Groth}\ \emph {et~al.}(2014)\citenamefont {Groth}, \citenamefont {Wimmer}, \citenamefont {Akhmerov},\ and\ \citenamefont {Waintal}}]{groth2014kwant}%
  \BibitemOpen
  \bibfield  {author} {\bibinfo {author} {\bibfnamefont {C.~W.}\ \bibnamefont {Groth}}, \bibinfo {author} {\bibfnamefont {M.}~\bibnamefont {Wimmer}}, \bibinfo {author} {\bibfnamefont {A.~R.}\ \bibnamefont {Akhmerov}},\ and\ \bibinfo {author} {\bibfnamefont {X.}~\bibnamefont {Waintal}},\ }\bibfield  {title} {\bibinfo {title} {{Kwant: a software package for quantum transport}},\ }\href {https://doi.org/10.1088/1367-2630/16/6/063065} {\bibfield  {journal} {\bibinfo  {journal} {New Journal of Physics}\ }\textbf {\bibinfo {volume} {16}},\ \bibinfo {pages} {063065} (\bibinfo {year} {2014})}\BibitemShut {NoStop}%
\bibitem [{\citenamefont {Datta}(1995)}]{datta1997electronic}%
  \BibitemOpen
  \bibfield  {author} {\bibinfo {author} {\bibfnamefont {S.}~\bibnamefont {Datta}},\ }\href {https://doi.org/10.1017/cbo9780511805776} {\emph {\bibinfo {title} {Electronic Transport in Mesoscopic Systems}}}\ (\bibinfo  {publisher} {Cambridge University Press},\ \bibinfo {year} {1995})\BibitemShut {NoStop}%
\bibitem [{\citenamefont {Xie}\ \emph {et~al.}(2022)\citenamefont {Xie}, \citenamefont {Chen}, \citenamefont {Xu},\ and\ \citenamefont {Liu}}]{xie2022effective}%
  \BibitemOpen
  \bibfield  {author} {\bibinfo {author} {\bibfnamefont {Y.}~\bibnamefont {Xie}}, \bibinfo {author} {\bibfnamefont {L.}~\bibnamefont {Chen}}, \bibinfo {author} {\bibfnamefont {J.}~\bibnamefont {Xu}},\ and\ \bibinfo {author} {\bibfnamefont {W.}~\bibnamefont {Liu}},\ }\bibfield  {title} {\bibinfo {title} {{Effective regulation of the electronic properties of a biphenylene network by hydrogenation and halogenation}},\ }\href {https://doi.org/10.1039/d2ra03673h} {\bibfield  {journal} {\bibinfo  {journal} {RSC Advances}\ }\textbf {\bibinfo {volume} {12}},\ \bibinfo {pages} {20088–20095} (\bibinfo {year} {2022})}\BibitemShut {NoStop}%
\bibitem [{\citenamefont {Cresti}\ \emph {et~al.}(2008)\citenamefont {Cresti}, \citenamefont {Nemec}, \citenamefont {Biel}, \citenamefont {Niebler}, \citenamefont {Triozon}, \citenamefont {Cuniberti},\ and\ \citenamefont {Roche}}]{cresti2008charge}%
  \BibitemOpen
  \bibfield  {author} {\bibinfo {author} {\bibfnamefont {A.}~\bibnamefont {Cresti}}, \bibinfo {author} {\bibfnamefont {N.}~\bibnamefont {Nemec}}, \bibinfo {author} {\bibfnamefont {B.}~\bibnamefont {Biel}}, \bibinfo {author} {\bibfnamefont {G.}~\bibnamefont {Niebler}}, \bibinfo {author} {\bibfnamefont {F.}~\bibnamefont {Triozon}}, \bibinfo {author} {\bibfnamefont {G.}~\bibnamefont {Cuniberti}},\ and\ \bibinfo {author} {\bibfnamefont {S.}~\bibnamefont {Roche}},\ }\bibfield  {title} {\bibinfo {title} {{Charge transport in disordered graphene-based low dimensional materials}},\ }\href {https://doi.org/10.1007/s12274-008-8043-2} {\bibfield  {journal} {\bibinfo  {journal} {Nano Research}\ }\textbf {\bibinfo {volume} {1}},\ \bibinfo {pages} {361–394} (\bibinfo {year} {2008})}\BibitemShut {NoStop}%
\bibitem [{\citenamefont {Mucciolo}\ \emph {et~al.}(2009)\citenamefont {Mucciolo}, \citenamefont {Castro~Neto},\ and\ \citenamefont {Lewenkopf}}]{mucciolo2009conductance}%
  \BibitemOpen
  \bibfield  {author} {\bibinfo {author} {\bibfnamefont {E.~R.}\ \bibnamefont {Mucciolo}}, \bibinfo {author} {\bibfnamefont {A.~H.}\ \bibnamefont {Castro~Neto}},\ and\ \bibinfo {author} {\bibfnamefont {C.~H.}\ \bibnamefont {Lewenkopf}},\ }\bibfield  {title} {\bibinfo {title} {{Conductance quantization and transport gaps in disordered graphene nanoribbons}},\ }\href {https://doi.org/10.1103/PhysRevB.79.075407} {\bibfield  {journal} {\bibinfo  {journal} {Phys. Rev. B}\ }\textbf {\bibinfo {volume} {79}},\ \bibinfo {pages} {075407} (\bibinfo {year} {2009})}\BibitemShut {NoStop}%
\end{thebibliography}%

\end{document}